\documentclass{iaus}
\usepackage{graphicx}

\newcounter{subfigure}

\def\Sec{\hbox{${}^{\prime\prime}$\llap{.}}}
 
\def\sec{\hbox{${}^{\prime\prime}$}}
\def\lae{\mathrel{<\kern-1.0em\lower0.9ex\hbox{$\sim$}}}
\def\gae{\mathrel{>\kern-1.0em\lower0.9ex\hbox{$\sim$}}}
\def\msun{M$_\odot$}
\def\ms{$\cal M - \sigma$}
\def\etal{et al.~}

\title[Cores, Nuclei and Supermassive Black Holes] 
{The Inner Workings of Early-Type Galaxies: Cores, Nuclei and
  Supermassive Black Holes}

\author[L. Ferrarese \& P. C\^ot\'e]   
{Laura Ferrarese$^1$ \and Patrick C\^ot\'e$^1$}

\affiliation{$^1$National Research Council of Canada, Herzberg 
Institute of Astrophysics, 5071 West Saanich Road, Victoria, 
BC, V9E 2E7, Canada \break
email: laura.ferrarese@nrc-cnrc.gc.ca; patrick.cote@nrc-cnrc.gc.ca\\
[\affilskip]}

\pubyear{2006}
\volume{238}  
\pagerange{001--999}
\date{??? and in revised form ???}
\jname{Black Holes: from Stars to Galaxies -- across the Range of Masses}
\editors{V. Karas \& G. Matt, eds.}
\begin{document}

\maketitle

\begin{abstract}
Recent years have seen dramatic progress in the study of the core and
nuclear properties of galaxies. The structure of the cores has been
shown to vary methodically with global and nuclear properties, as
cores respond to the mechanisms by which galaxies form/evolve. The
dynamical centers of galaxies have been found capable of hosting two
seemingly disparate objects: supermassive black holes (SBHs) and
compact stellar nuclei. In a drastic departure from previous beliefs,
it has been discovered that both structures are common: galaxies
lacking SBHs and/or stellar nuclei are the exception, rather than the
norm. This review explores the connection between cores, SBHs and
stellar nuclei in early-type galaxies, as revealed by the ACS Virgo
Cluster Survey.

\end{abstract}

\firstsection
\section{Introduction}

Cores, a term we will use loosely to describe the central few hundred
parsec region of a galaxy, represent an integral part in our
understanding of the global galactic structure, for very good reasons.
Cores act as recording devices of a galaxy history. Dynamical
timescales are shorter here than elsewhere in the galaxy; the
morphology, dynamics and history of star formation and chemical
enrichment of the cores are a sensitive tracer of the gas, dust and
dense stellar systems, either intrinsic or accreted through merging
events, that are drawn to the bottom of the potential well throughout
cosmic times. Furthermore, core and global properties are linked
through a number of scaling relations. In particular, those involving
supermassive black holes (SBHs) -- which are almost always associated
with galactic cores --  underscore the importance of nuclear feedback
in galaxy evolution (Ferrarese \& Merritt 2000; Gebhardt et al. 2000;
Graham et al. 2001; Ferrarese 2002; Haring \& Rix 2004).

The study of galactic cores received a tremendous push forward with
the deployment of the Hubble Space Telescope. HST images brought into
focus a plethora of structural features, including nuclear  stellar
disks, bars, ``evacuated'' regions (possibly scoured out by the
evolution of SBH binaries), and an entire spectrum of dust features -
from small irregular patches to large, organized dust disks. In
early-type galaxies, cores were found to fall in two distinct classes:
those with a shallow surface brightness profile, and those whose
surface brightness kept increasing, in roughly a power-law fashion, to
the innermost radius accessible given the resolution limit of the
instrument (Ferrarese et al. 1994; Lauer et al. 1995,2005;
Ravindranath et al. 1996; Rest et al. 2001).  Galaxies falling into
the first class have become known (somewhat unfortunately) as ``core''
galaxies, galaxies falling into the second class as ``power-law''. The
division between the two classes was found to correlate neatly with
galaxy luminosity, with core galaxies being  exclusively  bright giant
ellipticals, while fainter galaxies are (with few exceptions)
classified as power-laws. The stark separation between the two classes
has been attributed to differing  formation/evolutionary
histories. Power-law galaxies have been claimed to be the result of
dissipation during galaxy formation, with some authors further
claiming that all power-law galaxies host stellar disks; while
core-galaxies  are believed to be the result of the merging of fainter
(power-law) galaxies, and of their central SBHs.

Dynamical detections of SBHs exist in approximatively three dozen
galaxies (see Ferrarese \& Ford 2005 for a review); indeed, balancing
the SBH mass function from the QSO epoch to the present day requires
virtually all local galaxies brighter than a few $0.1L^*$ to host a
SBH (e.g. Shankar et al. 2004; Marconi et al. 2004). Recent
observations, however, have made it clear that SBHs are not the only
objects to enjoy a priviledged position at a galaxy's dynamical
center. Stellar nuclei, or nuclear star clusters,  have recently been
detected in a large fraction (70\% to 90\%) of galaxies of all Hubble
types and luminosities (B\"oker et al. 2002; Lotz et al. 2004; Grant
et al. 2005).  Follow-up spectroscopy of stellar nuclei in spiral
galaxies (Walcher et al. 2005,2006; Rossa et al. 2006) has shown them
to be massive, dense objects akin to compact star
clusters. Luminosity-weighted ages range from 10 Myr to 10 Gyr,
younger than the age of the galactic disk, and with the younger
clusters found preferentially in the later type spirals.

This review explores the connection between cores, nuclei and
supermassive black holes in light of recent results from the ACS Virgo
Cluster Survey (ACSVCS), the largest HST imaging survey designed
specifically to provide an unbiased characterization of the core
structure of early-type galaxies.

\section{The ACS Virgo Cluster Survey}
\label{sec:acsvcs}

The ACSVCS (C\^ot\'e et~al. 2004) consists of HST imaging for 100
members of the Virgo Cluster, supplemented by imaging and spectroscopy
from WFPC2, Chandra, Spitzer, Keck and KPNO.   The program galaxies
span a range of $\approx$ 460 in $B$-band luminosity and have
early-type morphologies: E, S0, dE, dE,N or dS0. All images were taken
with the Advanced Camera for Surveys (ACS; Ford et~al. 1998) using a
filter combination roughly equivalent to the $g$ and $z$ bands in the
SDSS photometric system. The images cover a $\approx$
200$^{\prime\prime}\times200^{\prime\prime}$ field with $\approx$
0\Sec1~ resolution ($\approx$8pc at the distance of Virgo, 16.5 Mpc).

This review summarizes results from the subset of ACSVCS papers which
deal with the morphology, isophotal parameters and surface brightness
profiles for early-type galaxies (Ferrarese et~al. 2006a), their
central nuclei (C\^ot\'e et~al. 2006) and scaling relations for nuclei
and SBHs (Ferrarese et~al. 2006b).  Other ACSVCS papers have discussed
the data reduction pipeline (Jord\'an et al. 2004a), the connection
between low-mass X-ray binaries and globular clusters (Jord\'an et al.
2004b), the measurement and calibration of surface brightness
fluctuation distances (Mei et~al. 2005ab), the connection between
globular clusters and ultra-compact dwarf galaxies (Ha\c{s}egan et
al. 2005), the luminosity function, color distributions and half-light
radii  of globular clusters (Jord\'an et al. 2006ab; Peng
et~al. 2006a), and diffuse star clusters (Peng et~al. 2006b).

\section{The Core Structure of Early-Type Galaxies}

Over the three-decade radial range between a few tens of parsecs and
several kiloparsecs (i.e. to the largest radii covered by the ACSVCS
images), the surface brightness profiles of the ACSVCS early-type
galaxies are well described by a simple S\'ersic model (S\'ersic 1968)
with index $n$ increasing steadily with galaxy luminosity. Notable,
and systematic, deviations from a S\'ersic model are however
registered in the innermost regions.  For eight of the 10 brightest
galaxies ($M_B \lesssim -20.3$) the measured inner profiles (typically
within 0\Sec5 to 2\Sec5, corresponding to 40 to 200pc) are shallower
than expected based on an inward extrapolation of the S\'ersic model
constrained by the region beyond.  For these galaxies,  the surface
brightness profile is best fitted by joining the outer S\'ersic
profile to an inner, shallower,  power-law component (such composite
models are referred to as ``core-S\'ersic'' Graham et~al. 2003;
Trujillo et~al. 2004),

The opposite is seen in fainter galaxies, $\approx$ 80\% of which show
a clear upturn, or inflection, in the surface brightness profile
within (typically) the innermost few tens of parsec region (see Figure
1 of P. C\^ot\'e, these proceedings). The upturn signals the presence
a stellar nucleus that is most likely structurally distinct from the
main body of the underlying galaxy. When a nucleus is present, the
inner surface brightness is, by definition,  larger than the inward
extrapolation of the outer S\'ersic model.

The picture that has emerged from the ACSVCS is therefore one in
which, in moving down the luminosity function from giant to dwarf
early-type galaxies, the innermost 100-parsec region undergoes a
systematic and smooth transition from light (mass) ``deficit''
(relative to the overall best fitting S\'ersic model) to light
``excess''. Although the subset of ACSVCS ``core-S\'ersic'' galaxies
coincides with the galaxies that were classified as ``cores'' in
previous investigations, there are critical differences between our
study and the ones that preceded it. Compared to previous work, the
ACSVCS has emphasized the role of stellar nuclei; the fact that the
frequency, luminosities and sizes of the ACSVCS nuclei are in
remarkable agreement  with those measured (using different techniques
and assumptions) by recent independent surveys in both early and late
type galaxies, supports the robustness of the ACSVCS analysis.
Recognizing the nuclei as separate components has allowed us to
revisit the issue of the division of early-type galaxies into ``core''
and ``power-law'' types. Such division was based on the fact that the
distribution of the logarithmic slopes, $\gamma = - d \log I / d \log
r$, of the inner surface brightness profile had been found to show
various degrees of bimodality.  Such bimodality is absent in the
$\gamma$ distribution of the ACSVCS galaxies. In agreement with
previous studies, in galaxies brighter than $M_B \approx -20.3$,
$\gamma$ is indeed found to decrease with galaxy luminosity, while the
opposite trend is seen for fainter galaxies, however, the transition
is smooth, rather than abrupt. In a further departure from previous
studies, we find that the low-$\gamma$ end of the distribution
(corresponding to the galaxies with the shallowest surface brightness
profiles) is occupied mostly by the faintest dwarf stystems, rather
than by the brightest giant ellipticals. We note here that the absence
of a bimodal behaviour in $\gamma$ does not automatically invalidate a
picture in which brighter galaxies evolve mainly through merging while
fainter systems are largely left untouched. Indeed, such picture does
not necessarily explain the perceived stark separation of galaxies in
``core'' and ``power-law'' types for which it was formulated. The
extent to which structural parameters are compromised by merging of
galaxies (and their supermassive black holes) depends on the the
masses of the progenitors (e.g., Bournaud et al. 2005; Milosavljevic
\& Merritt 2001); given a continuous distribution for the latter,
combined with a galaxy luminosity function heavily biased towards
low-mass systems, allows for the possibility of  a smooth transition
between progenitors and merger products.

\section{Compact Stellar Nuclei in the ACSVCS}

At the outset of the ACSVCS, it was known that at least $\approx$25\%
of the program galaxies contained nuclei, based on ground-based
classifications  from the VCC (Binggeli et~al. 1985). Stellar nuclei
in the ACSVCS images were identified by a variety of indicators,
including direct inspection of the ACS frames, color changes in the
$g-z$ color images, and sudden upturns in the surface brightness
profiles.  Based on these criteria, 60 to 80\% of ACSVCS galaxies host
stellar nuclei (with the precise fraction depending on galaxy
magnitude), in line with the fraction reported in both spiral and
elliptical galaxies based on recent high-resolution surveys (Carollo,
Stiavelli \& Mack 1998; Matthews et~al 1999; B\"oker et~al. 2002,
2004; Balcells \etal\ 2003; Lotz \etal\ 2004; Graham \& Guzman 2003;
Grant \etal\ 2005), but a factor $\sim 3$ higher than expected based
on the VCC.

Our analysis shows that surface brightness selection biases in the VCC
data are largely responsible for the difference:  in galaxies with
central $g$-band surface brightnesses lower than $\approx$
20.5~mag~arcsec$^{-2}$, the agreement between the ACSVCS and VCC is
nearly perfect, while above 19.5~mag~arcsec$^{-2}$, virtually all
nuclei were missed by the ground-based survey. Selection effects
might, of course, still be at work in the ACSVCS sample, implying that
our estimate for the frequency of nucleation, $f_n \approx 60-80\%$,
is almost certainly a lower limit on the true frequency.  As will be
discussed shortly, the luminosity and half-light radii of stellar
nuclei correlate strongly with the magnitude of the host galaxy; it is
therefore possible, for each galaxy classified as non-nucleated, to
determine whether a nucleus, if present, could have gone
undetected. Based on these tests, with very few exceptions, the only
galaxies for which the existence of a nucleus can be confidently
excluded are those brighter than $M_B \approx -20.3$ mag. These are
the same galaxies with central light ``deficits'' for which the
surface brightness profiles are well represented by ``core-S\'{e}rsic"
rather than S\'ersic models (\S 3)

\begin{figure}
\centering
\includegraphics[scale=0.6]{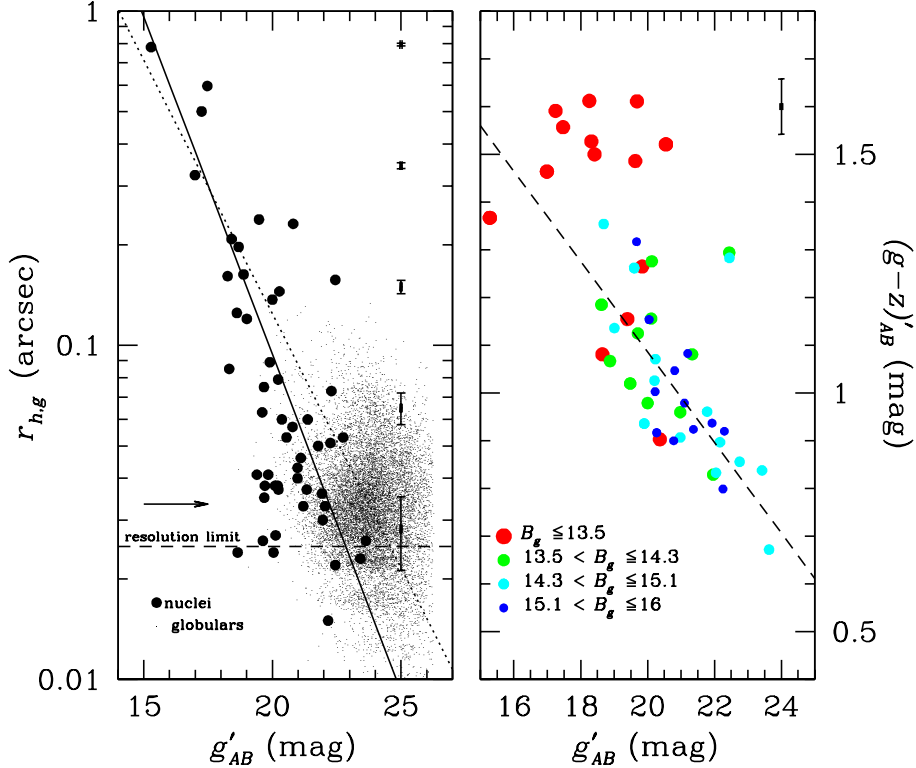}
\caption{{\it (Left Panel)} The size-magnitude relation, in the
$g-$band, for the 51 ACSVCS nuclei for which structural parameters
could be measured (solid circles) and the sample of globular clusters
from Jord\'an et al. (2006) (points). Typical errorbars for the
nuclear sizes are shown in the right hand side of the panel. The arrow
shows the "universal" half-light radius of 0\Sec033 ($\approx$ 2.7 pc)
for globular clusters in Virgo (Jord\'an et al. 2005), while the
dashed line shows a conservative estimate for the resolution limit of
the ACS images. The solid diagonal line shows the best fitting
relation for the nuclei ($r_h \propto L^{0.5}$), while the dotted line
shows the prediction of the globular cluster merging model of Bekki et
al. (2004).  {\it (Right Panel)} Color-magnitude diagram for the
ACSVCS nuclei.The size of the symbol for the nuclei is proportional to
the magnitude of the host galaxy, shown in the legend. The dashed line
shows the the best fit relation for the nuclei of galaxies fainter
than $B_T = 13.5$ mag.  
}
\label{fig:f1}       
\end{figure}

\subsection{Scaling Relations for Stellar Nuclei} 

For 51 galaxies in the ACSVCS the sharp upturn in the surface
brightness  within $\approx 1^{\prime\prime}$ is conspicuous enough
that a measurement of the nucleus' photometric and structural
parameters is possible. These parameters are recovered by adding a
central King model (King 1966) to the underlying S\'ersic component
when fitting the surface brightness profile.

The luminosity function of nuclei follows a Gaussian distribution with
dispersion in the range $1.5- 1.8$~mag and peak absolute $g-$band
magnitude $\approx -10.7$ mag, a factor $\approx 25\times$ brighter
than the peak of the globular cluster luminosity function. With a
half-dozen exceptions, nuclei in the ACSVCS galaxies are clearly
spatially resolved (thereby ruling out an AGN origin), with
individual sizes ranging from 62~pc down to the resolution limit of
2~pc, and a median half-light radius of $\langle r_h \rangle =
4.2$~pc.  Unlike globular clusters, for which size is largely
independent of magnitude, nuclear sizes are found to scale with
luminosity according to the relation $r_h \propto \cal
L$$^{0.50\pm0.03}$ (Figure 1, left panel).

One of the most credited models posits that the formation of nuclei
proceeds trough the coalescence of globular clusters drawn to the
bottom of the potential well by dynamical friction (e.g. Tremaine et
al. 1975). While the size-magnitude relation observed for the ACSVCS
nuclei is consistent with the prediction of such model (Bekki et
al. 2004), a more complex picture is put forth by the observations
that nuclei, again unlike globular clusters, display a color-magnitude
relation (Figure 1, right panel).  Monte Carlo simulations show that
mergers of globular clusters through dynamical friction are unable to
explain the observed color-magnitude relation; indeed the existence of
this relation suggests that the chemical enrichment of nuclei is
governed by local or internal factors, along the lines of the various
``gas accretion'' models  (e.g. Mihos \& Hernquist 1996). Note that
the nuclei's color-magnitude relation is better defined for galaxies
fainter than $M_B \approx -17.6$~mag, while the nuclei belonging to
brighter galaxies frequently show very red colors, $(g-z) \sim
1.5$. If confirmed (measurements are more uncertain for these nuclei,
due to the high underlying galaxy surface brightness), this
observation might suggest that these nuclei may constitute a separate
type of objects following a different formation route.

A third model, namely nuclear formation through two-body relaxation
around a black hole, is inconsistent with the observation that nuclei
are spatially resolved in most of the ACSVCS galaxies.  Nuclei formed
through this mechanism are predicted to extend to approximately 1/5 of
the SBH sphere of influence (e.g. Merritt \& Szell 2005), and would
therefore be  spatially unresolved by the ACS in all of the ACSVCS
galaxies, clearly contradicting our observations.

Finally, we note that the luminosity function and size distribution of
the ACSVCS nuclei shows remarkable agreement with those of the
``nuclear star clusters'' detected in spiral galaxies (B\"oker et
al. 2002,2004). This points to a formation mechanism for the nuclei
that is largely independent on both intrinsic and extrinsic factors,
such as host magnitude and Hubble type, and immediate environment.

\begin{figure}
\centering
\includegraphics[scale=0.7]{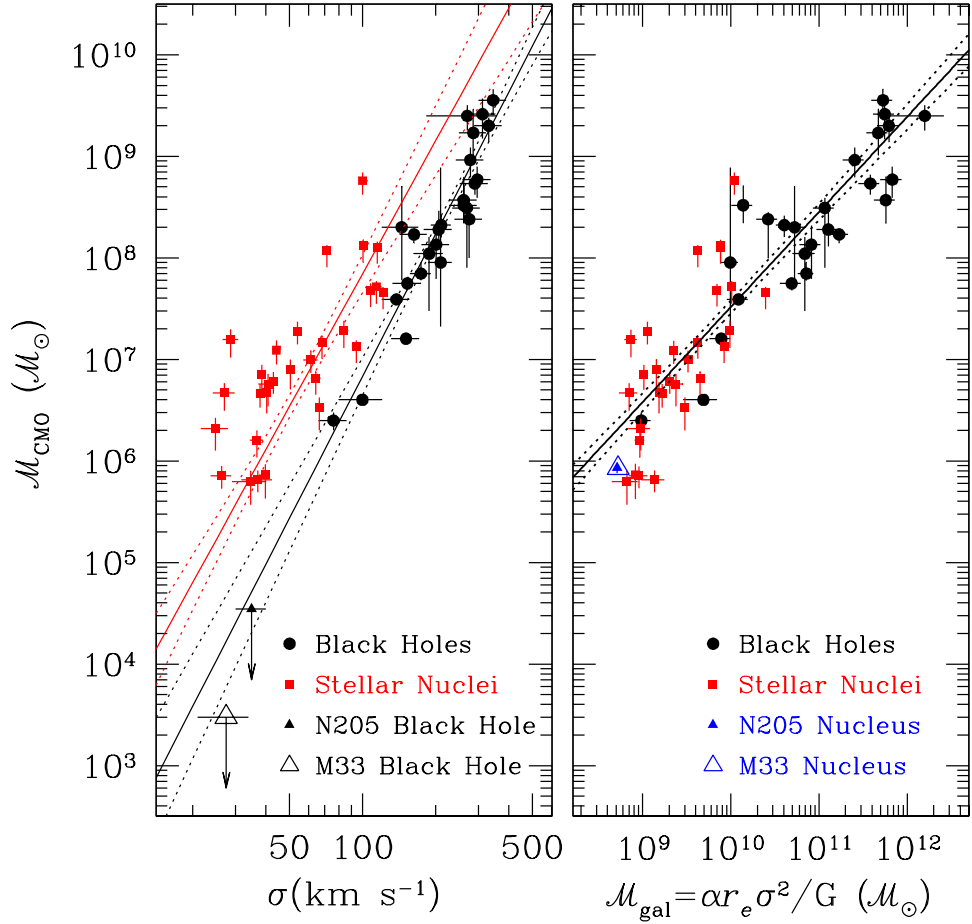}
\caption{{\it (Left Panel)} Mass of the CMO (stellar nuclei as red
squares and SBHs as black circles) plotted against the velocity
dispersion of the host galaxy. The solid red and black lines show the
best fits to the nuclei and  SBH samples, respectively, with 1$\sigma$
confidence levels shown by the dotted lines. {\it (Right Panel)} CMO
mass plotted against galaxy virial mass. The solid line is the fit
obtained for the combined nuclei and SBH sample.}
\label{fig:f2}       
\end{figure}

\vskip .2in
\section {Stellar Nuclei and Supermassive Black Holes}

The ubiquitousness of SBHs and stellar nuclei, and their unique
location at the dynamical centres of galaxies, are reasons to suspect
that a connection between the two might exist.

The ACSVCS data strongly support this view. The left panel of Figure 2
(from Ferrarese et al. 2006b) shows a recent characterization of the
relation between the masses of SBHs (black circles) and the stellar
velocity dispersion of the host bulge, originally discovered by
Ferrarese \& Merritt (2000) and Gebhardt et al. (2000). The
\ms~relation  is one of the tightest, and therefore most fundamental,
of the scaling relations for SBHs, and has been used extensively to
constrain the joint evolution of SBHs and galaxies (e.g., Haehnelt
2004 and references therein). The ACSVCS stellar nuclei (shown as red
squares) obey an \ms~relation with the same slope, although different
normalization, as the one defined by SBHs. This is a notable and
unexpected finding, suggesting close similarities in the formation and
evolutionary history of these two radically different structures
(McLaughlin et al. 2006). Furthermore, when $\sigma$ is combined with
the effective radius $r_e$ to produce a measure of the galaxy's virial
mass, $\cal M$$_{gal} \propto \sigma^2 r_e/G$, SBHs and stellar nuclei
are found to obey an identical  $\cal M$$  - \cal M$$_{gal}$  relation
(right panel of Figure 2;  see also C\^ot\'e et al. 2006). Remarkably,
the same relation is also found to hold in spiral galaxies (Rossa et
al. 2006) and to extend to dEs as faint as $M_B \approx -11.7$ mag
(Wehner \& Harris 2006).

These findings can be summarized as follows: a constant fraction,
$\cal M$$_{\rm CMO}/ \cal M$$_{gal} \approx 0.2$\%, of a galaxy total
mass is used in the formation of a nuclear structure, or ``central
massive object'' (CMO). This holds true for galaxies spanning a factor
$10^4$ \msun~in mass, all Hubble types, luminosities and
environments. In spite of their obviously different nature, SBHs and
stellar nuclei are nothing but complementary incarnations of CMOs -
they likely share a common formation mechanism and follow a similar
evolutionary path throughout their host galaxy's history. From the
prospective of a theoretical framework of galaxy evolution, the
commonalities between SBHs and stellar nuclei imply that both are
equally relevant: as the characterization of SBHs has been
instrumental in furthering our understanding of galaxy evolution (via
AGN feedback), so promises to be the characterization of stellar
nuclei (via superwinds and stellar feedback).

Several questions remain unanswered at this stage. Perhaps the most
intriguing is whether the formation of SBHs and stellar nuclei are
mutually exclusive. Nuclei are not present in the brightest ACSVCS
galaxies, the prototypical objects in which SBHs are expected to
reside, and for which a ``mass deficit'' has been attributed to the
evolution of supermassive black hole binaries (Milosavjevic \& Merritt
2001). At the other extreme of the luminosity range spanned by the
ACSVCS galaxies, NGC205 and M33, for which there is no evidence of a
SBH (Merritt \etal\ 2001; Gebhardt \etal\ 2001; Valluri \etal\ 2003),
host stellar nuclei that follow the same scaling relations as the
nuclei detected in the ACSVCS galaxies (Figure 2, right panel).  It is
possible that nuclei form in every galaxy, but are subsequently
destroyed in the brighest system as a consequence of the evolution of
SBH binaries. Alternatively, it is possible that collapse to a SBH
takes place preferentially in the brightest galaxies, while in fainter
systems, the formation of a stellar nucleus is the most common
outcome. In the latter case, nuclei could represent ``failed black
holes'', low-mass counterparts of the SBHs detected in the brightest
galaxies.

The ACSVCS collaboration is currently persuing several programs aimed
at studying the dynamics and stellar population of the ACSVCS galaxies
and nuclei; a similar investigation is underway for a sample of 43
early-type galaxies in the Fornax Cluster (Jord\'an et al. 2006).
These projects promise to shed further light on the core structure  of
early-type galaxies, their nuclei and their inter-relation to SBHs.

\begin{acknowledgments}
More rewarding than the results themselves has been to work with a
great team. The friendship,  hard work and dedication of each member
of the ACSVCS team is very gratefully acknowledged.
\end{acknowledgments}

\newpage

\setcounter{page}{001}

\title[A Critical Comparison of Nuker and Core-S\'ersic/S\'ersic Models.]{}

\author[L. Ferrarese et al.]{}

\maketitle
\begin{centering}
\vskip 100pt 
\normalfont\LARGE\fontswitch\bfseries{Parametric Representation of Surface Brightness Profiles: a Critical Comparison of Nuker and Core-S\'ersic/S\'ersic Models.}\par
\vskip 10pt 
\normalfont\large\fontswitch\bfseries\baselineskip=12pt
Laura Ferrarese$^1$,
Patrick C\^ot\'e$^1$,
John P. Blakeslee$^2$,
Simona Mei$^3$,
David Merritt$^4$,
\and Michael J. West$^{5,6}$
\vskip 4pt 
\normalfont\small 
$^{1}${Herzberg Institute of Astrophysics, National
Research Council of Canada, 5071 West Saanich Road, Victoria, BC,
V8X 4M6, Canada}\\
$^{2}${Department of Physics, Washington State University,
Webster Hall 1245, Pullman, WA 99164-2814}\\
$^{3}${GEPI, Observatoire de Paris, Section de Meudon, 5 Place J.Janssen, 92195 Meudon Cedex, France}\\
$^{4}${Department of Physics, Rochester Institute of
Technology, 84 Lomb Memorial Drive, Rochester, NY 14623}\\
$^{5}${Department of Physics \& Astronomy, University of Hawai'i,
Hilo, HI 96720}\\
$^{6}${Gemini Observatory, Casilla 603, La Serena, Chile}\\
\par
\vskip 8pt
\end{centering}

\begin{abstract}
The parameterization of the surface brightness profiles of early-type
galaxies has been instrumental in characterizing scaling relations and
in defining the properties of these systems. In the study of the core
properties (i.e. within the innermost few hundred parsecs), the  most
commonly used parameterization is given by the so called ``Nuker"
model (Lauer et al. 1995), described by an inner and outer power law
joined at a ``break" radius. In recent years, however, shortcoming of
the Nuker model have started to become apparent (e.g. Graham et
al. 2003). Indeed, Ferrarese et al. (2006) and C\^ot\'e et al. (2006)
found it necessary to adopt a different parameterization in their
analysis of the surface brightness profiles of a sample of 100
early-type galaxies observed with the HST Advanced Camera for Surveys
as part of the ACS Virgo Cluster Survey (ACSVCS). In the ACSVCS
analysis, core-S\'ersic or S\'ersic models are claimed to provide good
descriptions of the surface brightness profiles from parsec to
kiloparsec scales, and are adopted in defining the properties of
compact stellar nuclei. In this contribution, we present a more
detailed comparison of Nuker and core-S\'ersic/S\'ersic models. This
comparison is motivated by a recent astro-ph posting (Lauer et al.,
astro-ph/0609762) where, based on HST/WFPC1 or WFPC2 images of  22 of
the ACSVCS galaxies, it is  argued that the S\'ersic and core-S\'ersic
models presented by Ferrarese et al.\ (2006) provide inadequate fits
to the surface brightness profiles and that such models lead to the
identification of spurious nuclear features. We show that the  Lauer
et al.\ criticisms are based on faulty assumptions and  misrepresent
the ACSVCS analysis. We further show that the Nuker model
parameterization used by Lauer et al.\ not only fails to reproduce the
surface brightness profiles on kiloparsec scales, but is also not a
particularly good representation of the profiles of the ACSVCS
galaxies on parsec scales. Indeed, we argue that, for several of the
galaxies in common with the ACSVCS sample, the Nuker model fits of
Lauer et al.\ were likely biased by the lower signal-to-noise ratio
and limited spatial extent of the WFPC1 or WFPC2 data used in their
analysis. These shortcomings are probably responsible for the fact
that Lauer et al.\ failed to recognize and characterize the properties
of stellar nuclei in many early-type galaxies.
\end{abstract}

\setcounter{section}0
\section{Introduction}

The ACS Virgo Cluster Survey (ACSVCS) is an HST project designed to
study the globular cluster systems and core properties of early-type
galaxies. The survey employed the Wide Field Channel (WFC) of the
Advanced Camera for Surveys (ACS) to image, in the $g$ and $z$ bands,
a representative sample of early-type (E, S0, dE, dE,N, S0 and S0,N)
confirmed members of the Virgo Cluster. The galaxies span a factor 460
in $B$-band luminosity and, at their nearly identical distance of 16.5
Mpc, are observed at the same high spatial resolution of $6.7$ pc. In
terms of sample size, completeness in both luminosity and
morphological type, spatial resolution, radial coverage and data
homogeneity, the ACSVCS represents the best sample of early-type
galaxies observed with HST to date.

The core and global properties of the ACSVCS galaxies have been
characterized in Ferrarese et al.\ (2006a, hereafter F06) and C\^ot\'e
et al.\ (2006, hereafter C06). In these papers, it was argued that
S\'ersic (S\'ersic 1968) or core-S\'ersic models (Graham et al.\ 2003)
provide a superior description of the surface brightness profiles of
early-type galaxies than the more commonly adopted ``Nuker" model
(Lauer et al.\ 1995), as this latter model fails to capture the
curvature of galaxy brightness profiles on kiloparsec scales. It was
further shown in the ACSVCS that the actual profiles vary in a smooth
and predictable fashion as one moves down the luminosity function for
early-type galaxies, as shown in Figure 1. Relative to the inward
extrapolation of the S\'ersic model that best fits the surface
brightness profiles between a few tens to several thousands of
parsecs, bright galaxies show a luminosity ``deficit": i.e., their
profiles are shallower and have lower surface brightness than expected
based on a S\'ersic characterization. Fainter galaxies, on the
contrary, show luminosity ``excesses" within the inner few parsecs relative
to the S\'ersic laws that best fit the profiles on larger scales. Such
excesses are identified with stellar nuclei, and are often associated
with a change in the $(g-z)$ color, an inflection in the surface
brightness profile, and a change in the isophotal parameters
(ellipticity and position angle) relative to the main body of the
galaxy. When the main body of the galaxy is considered (i.e.,
excluding a nuclear component), the distribution of inner profile
slopes, $\gamma = -d\log I / d\log r$, is found to be unimodal with
the low-$\gamma$ peak (corresponding to the shallowest profiles)
occupied by the {\it faintest} galaxies in the sample. In defining
$\gamma$, $I$ is the intrinsic (prior to PSF convolution) intensity of
the core-S\'ersic or S\'ersic model that best fits the global profile,
and the derivative is measured at 0\Sec1.

\setcounter{figure}{0}
\begin{figure}
\centering
\includegraphics[scale=0.6]{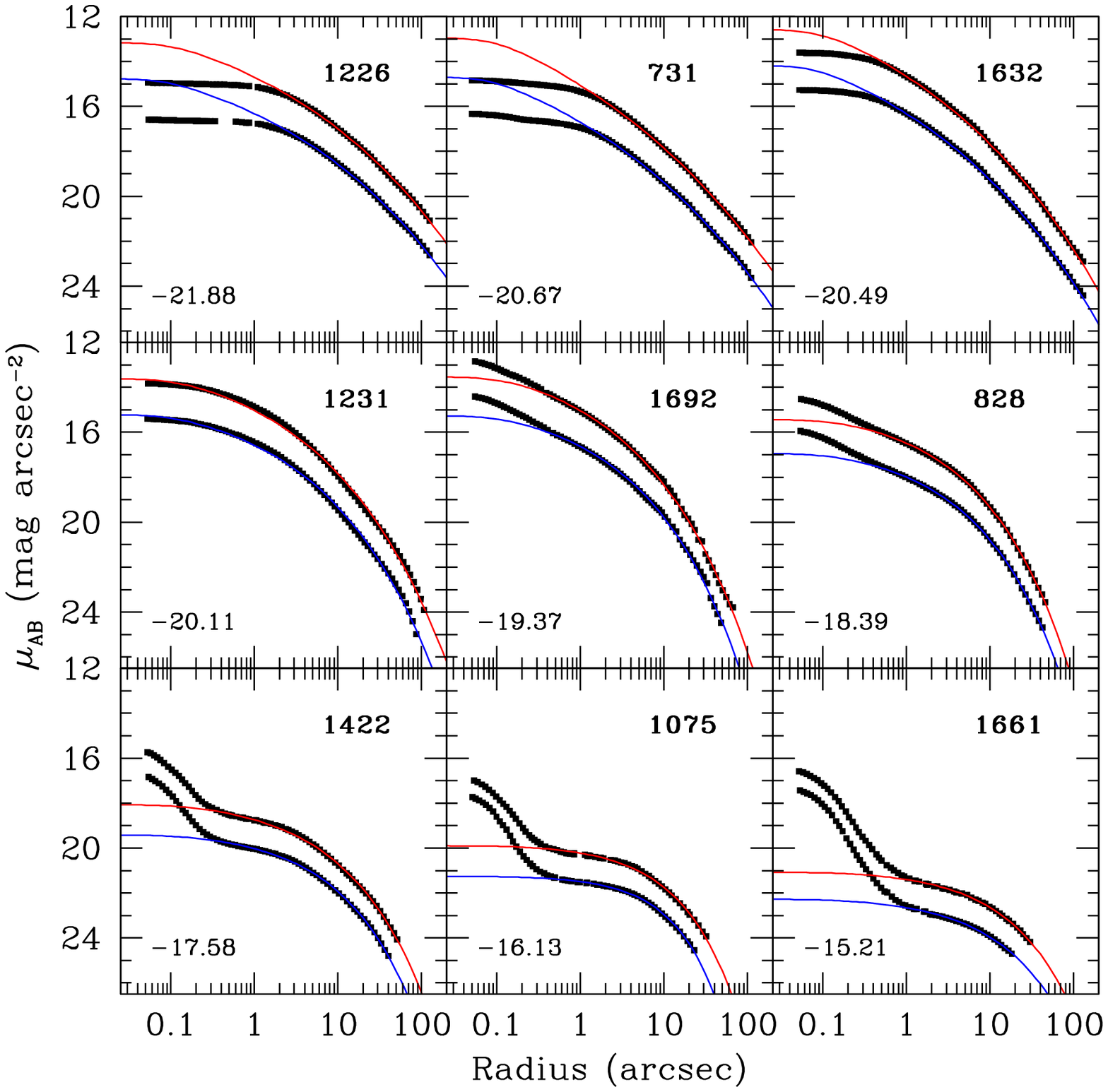}
\caption{Representative surface brightness profiles for nine
early-type galaxies from the ACSVCS. The galaxies span a factor of
$\sim 460$ in  $B-$band luminosity -- the $B-$band magnitude is listed
in the bottom left of each panel. For each galaxy, we show the
azimuthally-averaged brightness profile in the $g$ and $z$ bands
(lower and upper profiles, respectively). The solid curves show
S\'ersic models fitted to the profiles beyond $\sim
0\Sec2$-$2$\sec. Note the gradual progression from a central light
``deficit" to ``excess", with a transition at $M_B \sim -20$ (see
Ferrarese et~al. 2006a and C\^ot\'e et~al. 2006 for details).
}
\end{figure}

The analysis and interpretation presented in F06 and C06 have been
criticized in a recent astro-ph posting (Lauer, Gebhardt, Faber,
Richstone, Tremaine, Kormendy, Aller, Bender, Dressler, Filippenko,
Green \& Ho 2006, astro-ph/0609762).  The Lauer et al.\ posting is
divided into two parts. In the first four sections, the authors
address the issue of bimodality in the distribution of the logarithmic
slope, $\gamma = -d\log I / d\log r$, of the inner surface brightness
profiles of early type galaxies. To this end, Lauer et al.\ compile
$\gamma$ values from the literature. However, galaxies were included
only if $\gamma$ was derived, in the original investigation, by
fitting a Nuker model to the surface brightness profile measured from
HST images. A total of 219 galaxies satisfy this criterion; these
galaxies were observed as part of five independent projects, each
employing a different instrument and/or filter combination, and
analyzed by five independent groups, often with different
methodologies  (Lauer et al.\ 1995, 2005; Rest et al.\ 2001; Laine et
al.\ 2002; Quillen et al.\ 2000; Ravindranath et al.\ 2001).  Galaxies
which were observed with HST, but for which Nuker model fits are not
available in the literature, were simply excluded in the Lauer et
al. analysis.

Based on their compilation, Lauer et al.\ find the distribution of
$\gamma$ to be bimodal, with the low $\gamma$ peak (corresponding to
shallow surface brightness profiles) occupied exclusively by the {\it
brightest} galaxies in the sample, in contrast to the findings of
F06. In the second part of their astro-ph posting, Lauer et al.\
investigate the cause of such discrepancy and claim that the results
of F06 are invalidated by an inadequate analysis of the
data. Specifically, Lauer et al.\ claim that: (1) the S\'ersic and
core-S\'ersic models used by F06 to fit the profiles do a poor job at
describing the data at small radii, thus leading to an incorrect
measurement of $\gamma$; and (2) the inherent inadequacy of the
S\'ersic models fitted to most galaxies forces F06 to introduce ``{\it
ad hoc} stellar nuclei'', exacerbating a bias in $\gamma$.

The issue of bimodality (or lack thereof) in the distribution of inner
profile slopes will be addressed in several forthcoming papers. The
present contribution focuses on the second part of the Lauer et al.\
posting, and is intended to correct factual errors, misleading and
incorrect statements, and logical inconsistencies made in Lauer et
al..  In order to better understand what follows, it is useful to
clarify some of the differences in the  approach followed by Lauer et
al.\ and F06:

\begin{itemize}

\item{{\it Sample selection.} The samples used by Lauer et al.\ and
the ACSVCS sample differ dramatically in their selection function. No
objective selection criteria describe the former, the only commonality
between the  Lauer et al.\ galaxies being the fact that they were
observed with HST and that Nuker model parameters were available in
the literature. The sample used by F06, on the other hand, is a
representative (nearly magnitude limited) sample of 100 early-type
galaxies located at a common distance of $\approx 16.5$ Mpc, and
observed with an identical observational set-up (HST/ACS F475W and
F850LP). There are 27 galaxies in common between Lauer et al.\  and
F06; however, in the compilation of Lauer et al.\ the surface
brightness profiles for these galaxies were measured from WFPC1/F555W
images (9 galaxies), WFPC1/F785LP images (1 galaxy), WFPC2/F555W
images (12 galaxies), WFPC2/F702W images (3 galaxies) NICMOS2/F160W
images (1 galaxy) and NIC3/F160W images (1 galaxy).  Issues related to
the sample selection will be discussed in more detail in \S 2.}

\item{{\it Choice of parametrization of the surface brightness
profile}. For the galaxies included in Lauer et al., the profile was
parametrized using a Nuker model. F06 adopt a core-S\'ersic model
(Graham et al.\ 2003; Trujillo et al.\ 2004) for galaxies brighter
than $M_B \approx -20.3$ ($\sim 10$\% of the sample), and a S\'ersic
model (S\'ersic 1968) for fainter galaxies.\footnote{In what follows,
for convenience we will refer to these models as the ``ACSVCS
models".} When a stellar nucleus is present, as is the case for most
galaxies fainter than $M_B \approx -20.3$, it is described as a
PSF-convolved King model, added to (and fitted at the same time as)
the S\'ersic model representing the galaxy profile beyond the nuclear
component (i.e., for radii larger than a few 0\Sec1). There are
several reasons why Nuker models were not adopted for the analysis of
the ACSVCS data. Nuker model parameters have been shown to depend on
the radial extent of the data that is being fitted (Graham et al.\
2003), to the point that their physical interpretation is problematic
(this is a particular concern in the case of a sample of galaxies at
different distances, as in the compilation of Lauer et
al.). Furthermore, Nuker models asymptote to power-laws on large
scales, while real galaxies almost universally exhibit continuous
curvature on a log-log plot. This critical point is lost for the
galaxies in the Lauer et al.\ compilation, due to the small radial
extent of the data. However, the ACS data analyzed by F06 (and
supplemented, for the brightest galaxies, with ground based data)
encompass the curvature of the profiles at kiloparsec scales,
rendering the Nuker model a completely inadequate choice of
parametrization.}

\item {\it Treatment of the data.} For the galaxies in the Lauer et
al.\ compilation, Nuker models are fitted to deconvolved data (except
for the galaxies drawn from the Ravindranath et al.\ 2001 sample, for
which convolved models were fitted to data in the observational
plane). F06, on the other hand, fit PSF-convolved models to ACS data
in the observational plane. While there are pros and cons to both
methodologies, the Lauer et al.\ claim that ``recognizing subtle
systematic failures of the models is considerably more difficult in
the observed domain'' is entirely unsubstantiated. Deconvolution of
data is an inherently ill-posed process, and the instability is
greater the lower the signal-to-noise (S/N) ratio of the data, or the
larger the PSF compared to intrinsic physical scales (e.g. Craig and
Brown 1986). Results can depend critically on the choice of
regularization scheme, an issue not discussed by Lauer et al., and
this is most true near the center where the intrinsic profiles vary
rapidly on the scale of the PSF. Convolution of a noiseless model, on
the other hand, is a well-posed mathematical operation with a unique
solution.  Deconvolution is appropriate when attempting to
characterize the data in a non-parametric way (e.g. Merritt \&
Tremblay 1994), but when the goal is to compare data with empirical
models, it is always more appropriate to project the model into the
observed plane than vice versa (e.g. King 1975).

\item{{\it Identification of stellar nuclei.} Lauer et al.\ identify
nuclei as upturns relative to the best fitting Nuker model. F06
identify nuclei as upturns relative to the best fitting S\'ersic
model, and from a variety of other indicators, including visual
inspection of the images, and sudden changes in the isophotal
parameters and color profiles within the inner few 0\Sec1. As will be
shown in \S 4, identification of stellar nuclei is less ambiguous in
the case of the ACSVCS data, rather than the WFPC1 or WFPC2 data used
by Lauer et al.\ This is due to the higher S/N, larger radial extent,
and (compared to the WFPC1 data) higher resolution afforded by the
ACS, as well as to the availability of color images.}

{\item {\it Definition of $\gamma$.} Lauer et al.\ measure $\gamma$ at
either the resolution limit of the instrument (judged to be between
0\Sec02 and 0\Sec1) or the innermost radius that they deem to be
unaffected by a nuclear component, whichever is largest. Although
Lauer et al.\ do not tabulate the radii at which $\gamma$ is measured
for each galaxy, their figures indicate that these radii vary by at
least a factor 10, from 0\Sec02 to over 0\Sec2. By contrast, F06
measure $\gamma$ at a consistent angular scale of 0\Sec1. Because of
the common distance of the ACSVCS galaxies, this angular scale
corresponds to the same physical scale, $\sim 8$~pc, for all
galaxies. If a nucleus is present, the slope is measured from the
inward extrapolation of the S\'ersic model best fitting the profile in
the region unaffected by the nucleus (generally a few 0\Sec1 to
$\lesssim$ 100\sec). }

\end{itemize}

\section{Sample Comparison}

A comparison of the Lauer et al.\ and the ACSVCS samples is given in
Table 1.  As mentioned above, the Lauer et al. analysis is based on a
compilation of published Nuker model parameters fitted to surface
brightness profiles of early-type galaxies by a variety of groups
(Lauer et al.\ 1995,2005; Rest et al.\ 2001; Laine et al.\ 2002;
Quillen et al.\ 2000; Ravinandranath et al.\ 2001). The only
commonality between their galaxies is that all were observed with HST
(albeit with different instruments and filters) and all were fitted
using a Nuker model. Almost all galaxies have early-type morphologies,
although a handful of spiral bulges are included. HST data for which a
Nuker model fit was not available in the literature were excluded from
the outset.

\begin{table*}[t] 
\scriptsize
\caption{Comparison of the Lauer et al.\ and ACSVCS Samples}
\bigskip
\begin{tabular}{lccccl} 
Instrument  &
No. of &
Range in &
Range in &
Mean Spatial &
Source \\
&
Galaxies &
Distance (Mpc)&
$M_B$ (mag) &
Resolution (pc) &
\\
\hline
 & & & & & \\
\multicolumn{6}{c}{Lauer et al.\ 2006}\\
 & & & & & \\
WFPC2/F555W &               63    &    10.2 to 92.2   &   $-$18.0   to   $-$23.6 &  8.2 &  Lauer et al.\ 2005\\
WFPC2/F702W &               46    &    13.4 to 50.1   &   $-$18.4   to   $-$22.3 & 10.6 & Rest et al.\ 2001\\
WFPC2/F814W &               60    &    38.2 to 209    &   $-$21.8  to    $-$25.3  & 49.7 & Laine et al.\ 2002\\
WFPC1/F555W &               26    &    13.3 to 321    &   $-$15.6   to   $-$23.8 &  27.1 & Lauer et al.\ 1995\\
            &                     &                 &                          &        & Faber et al.\ 1997\\
NIC2/F160W &                ~9    &    12.9 to 73.3   &   $-$18.6   to   $-$23.3 & 22.9 & Quillen et al.\ 2000\\
NIC2/F160W or &            15    &    \phantom{1}3.5 to 56.0   &   $-$17.2 to $-$22.9 & 15.2 &  Ravindranath et al.\ 2001\\
NIC3/F160W &                      &                 &   &   &  \\
\hline
 & & & & & \\
\multicolumn{6}{c}{Ferrarese et al.\ (2006a) (ACSVCS)}\\
 & & & & & \\
ACS/F475W and      & 100     &  16.5 & $-15.1$ to $-21.8$  & 6.7 & C\^ot\'e et al.\ 2004  \\
F850LP &                      &                 &   &   &  \\
\hline
\end{tabular}
\end{table*}

The Lauer et al.\ sample therefore consists of  a not-easily
characterizable mix of parameters measured by five independent groups,
most, but not all, from deconvolved profiles of galaxies spanning a
factor 100 in distance (Figure 2), observed with four different
instruments and four different filters (from the $V$ to $H$ bands),
spanning a factor five in angular resolution (from 0\Sec02 to 0\Sec1
according to the authors), and a factor 65 in spatial resolution (from
2.4 pc to 156 pc, Figure 3).

\begin{figure}
\includegraphics[scale=0.7]{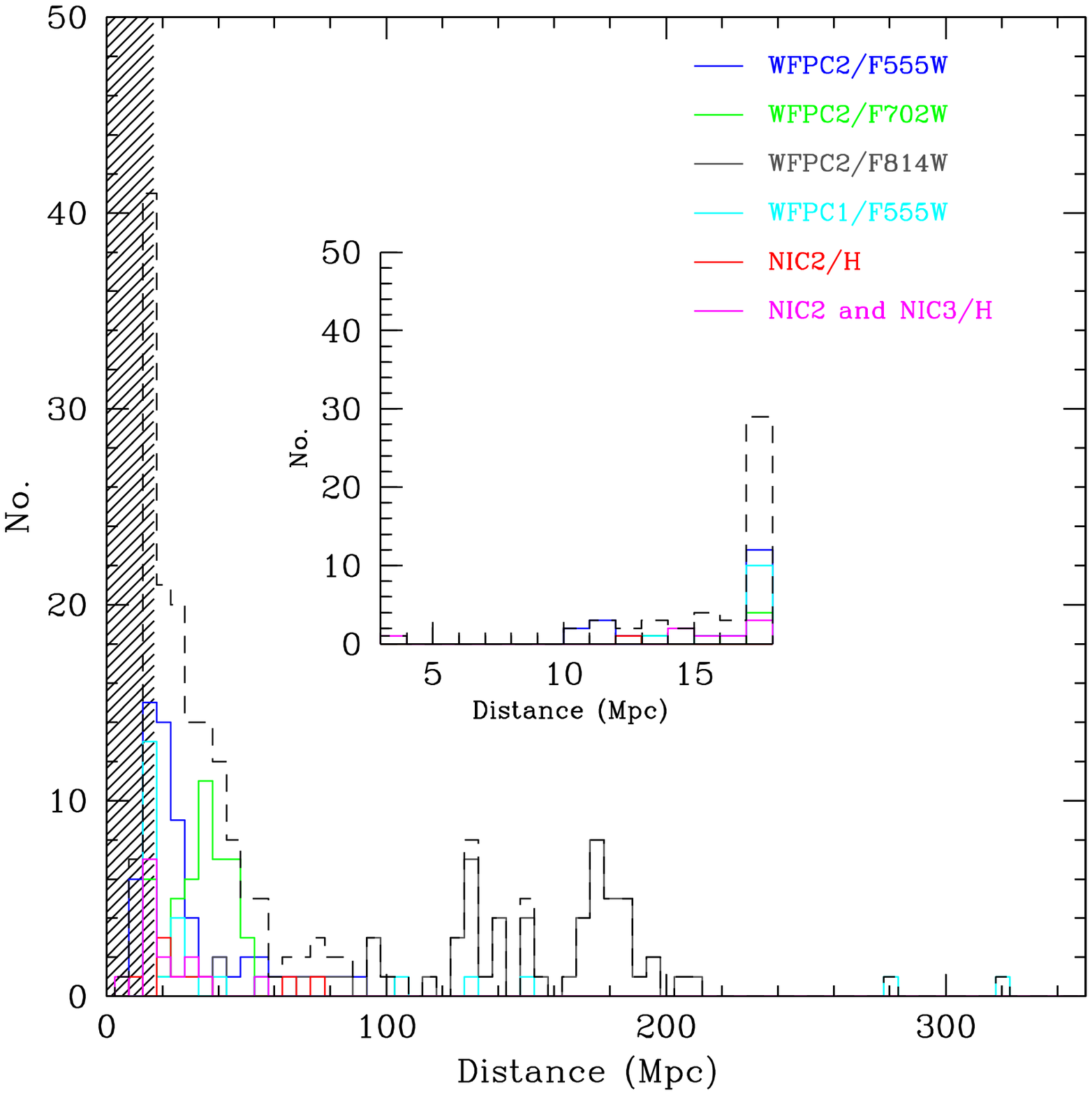}
\caption{The distribution of distances for the galaxies included in
the Lauer et al.\ posting. The dashed region identifies galaxies 
at the same distance as, or closer than, the ACSVCS galaxies 
(accounting for the fact that Lauer et al. place the
mean distance of Virgo at 17.9 Mpc) - the same region is expanded in the
inset. The dashed black curve shows the cumulative distribution.  }
\end{figure}

\begin{figure}
\includegraphics[scale=0.7]{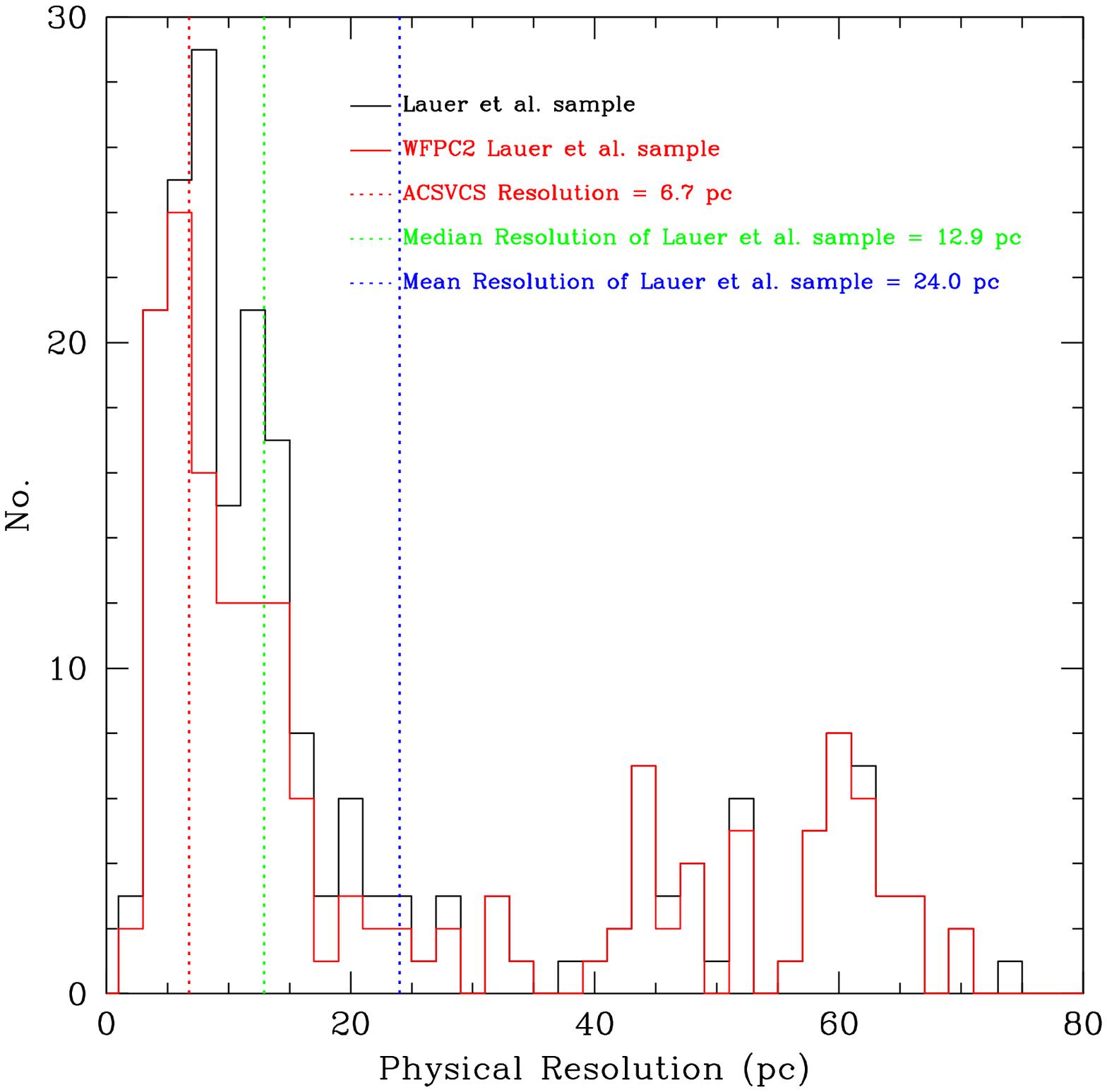}
\caption{The distribution of physical resolution (calculated based on
the FWHM of the PSF) for the 219 galaxies that comprise the Lauer et
al.\ sample. The ACSVCS resolution, 6.7 pc, is significantly higher
than the mean or median resolution of the Lauer et al.\ sample. If
only data observed with WFPC2 are considered (for a total of 169
galaxies), the mean resolution is 23.6 pc and the median resolution is
11.9 pc, in both cases lower than that of the ACSVCS sample, as stated
in F06. Two additional Lauer et al.\ galaxies (not shown) have
resolutions of 135 pc (A1020) and 155 pc (A1831) }
\end{figure}

Comparing the sample used by Lauer et al.\ to that from the ACSVCS
(Table 1), it is immediately evident that the ACSVCS sample: (1) is
larger than any single sample previously observed with an indentical
instrument/filter configuration; (2) has superior mean spatial
resolution;  (3) is by far the most homogeneous in terms of
environment/distance (all galaxies being located in a single cluster
and at approximatively at the same distance);  (4) is the only sample
containing a large number of low- and intermediate-luminosity galaxies
($\approx 60$ galaxies with $M_B \gtrsim -18$); and (5) is the only
sample for which color information is available for all galaxies.

Remarkably, Lauer et al.\ claim that the ACSVCS sample offers no
resolution advantage compared to existing HST samples and, in several
instances, assert that the data used in their analysis has better
angular and comparable spatial resolution as the ACS data used by F06:
e.g., (1) ``While the VCS ACS/WFC images have lower angular resolution
than the WFPC2/PC F555W images used for much of [our]
sample....''\footnotemark; and (2) ``A comparison of WFPC2/PC1 and
ACS/WFC PSFs shows that WFPC2 actually provides significantly better
angular resolution.... The present sample has 49 galaxies at Virgo
distance or closer, and a substantial number that are no more than
50\% more distant; both the present and [ACS]VCS samples are probing
the same physical scales in the galaxies.''

\footnotetext{In fact, slightly more than a quarter of the galaxies in
the Lauer et al. compilation was observed with HST/WFPC2/F555W.}

These statements are misleading. While it is true that 49 galaxies in
the Lauer et al.\ sample\footnotemark~are at the distance of Virgo (28
galaxies) or closer (21 galaxies), only 27 of these were observed with
WFPC2, and of these six were observed with F702W, which provides a
lower resolution than F555W. Thirteen of the 49 galaxies were observed
with WFPC1, and 9 with NICMOS, instruments that both have
significantly lower resolution than ACS/WFC. Figure 3 shows the
distribution in physical resolution (FWHM) for the complete sample
used by Lauer et al.\ (in black) and the subset of WFPC2 data (in
red). The resolution of the ACSVCS data is 6.7 pc (for all
galaxies). The mean resolution of the Lauer et al.\ complete sample is
24.0 pc, and the median resolution is 12.9 pc; the same numbers for
the subset of the Lauer et al.\ sample that used WFPC2 data are 23.6
pc and 11.9 pc. The ACSVCS spatial resolution is therefore between 2
and 3.5 times better, as already stated in F06. There is no question
that on the whole it is the ACSVCS data, not the Lauer et al.\ sample,
that provides a sharper (and more homogeneous) view of the innermost
regions of early-type galaxies.

\footnotetext{For simplicity, we will henceforth refer to the
WFPC1/WFPC2/NICMOS data that were used to fit the Nuker model
parameters compiled by Lauer et al.\ (Table 1) as the ``Lauer et al.\
data" although much of these data were taken by teams unrelated to
Lauer and collaborators.}

\section{Nuker vs. S\'ersic Models}

\subsection{Data Presentation in Lauer et al.}

In their Figures 11 and 12, Lauer et al.\ show surface brightness
profiles from deconvolved WFPC2/F555W or WFPC1/F555W data (with the
exception of NGC4464 = VCC1178, for which ACS data are shown). The
blue curves in their figures show the Nuker model that was designed to
best fit the WFPC1/WFPC2 surface brightness profile given as a
function of semi-major axis length. The red line shows the intrinsic
(prior to PSF convolution) ACSVCS model, with parameters given in
F06. Note that the ACSVCS models were fitted (after PSF convolution)
to ACS F475W data in the observational plane and cast as a function of
mean geometric radius. To ``account" for the filter and x-axis
mismatch between the ACSVCS models and the WFPC1/WFPC2 data against
which Lauer et al.\ plot those models, Lauer et al.\ scale the data
vertically by a constant factor and multiply the semi-major axis
length by a factor involving ellipticity (presumably also derived from
the WFPC1/WFPC2 F555W data, although this is not stated explicitly) to
convert it to mean geometric radius. No measures were taken to match
the ACSVCS models to the PSF of the deconvolved WFPC1/WFPC2 images.

Lauer et al.\ argue that such a comparison is legitimate on the basis
of the fact that profiles from deconvolved WFPC1, WFPC2 and ACS data
agree. But this point is irrelevant: even neglecting differences in
the filter and radial definition between the ACSVCS models and the
data against which they are plotted, the PSF of a deconvolved image is
not a $\delta$-function, and does not match the PSF of the ACSVCS
model prior to convolution. A correct analysis must compare the
performance of each model against the data used to fit those models.

\subsection {A Fair Comparison}

Figures 4a to 9a correspond to Figures 11, 12 and 13 of Lauer et al.\
In our Figures, however, each model is shown against the data used to
fit that model: the Nuker models used by Lauer et al.\ are plotted in
blue against the deconvolved WFPC1 or WFPC2 data used to perform the
fit (shown as a function of semi-major axis length), while the red
curves represent the PSF-convolved ACSVCS core-S\'ersic or S\'ersic
models overplotted on the ACS/WFC F475W data in the observational
plane (shown as a function of mean geometric radius). The first
crucial point to note is the much larger spatial coverage of the
ACSVCS compared to the WFPC2 or WFPC1 data: thanks to the larger FOV
of the ACS, the curvature of the profile on kiloparsec scales can now
be appreciated. It is also immediately apparent that such curvature
cannot be reproduced by the power-law behavior of the Nuker model on
large scales. Unlike the Nuker models, the ACSVCS models do a
remarkably good job at fitting the entire profile of the galaxies, in
many cases over more than three decades in radius. The systematic
failure of the Nuker model on large scales was not evident in the
figures shows in the Lauer et al. posting, where profiles are not
plotted on scales larger than 1-10\sec.

\renewcommand{\thefigure}{\arabic{figure}\alph{subfigure}}
\setcounter{subfigure}{1}
\begin{figure}
\includegraphics[scale=0.7]{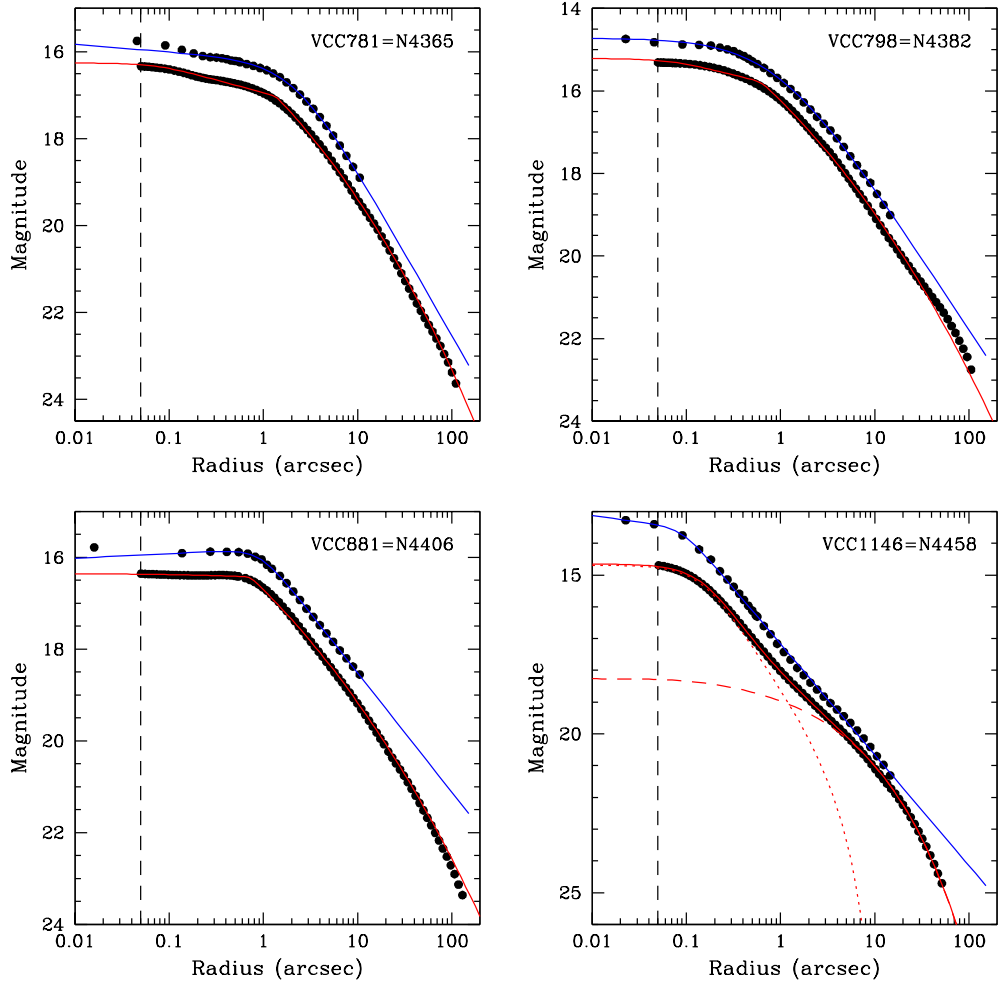}
\caption{The equivalent of the first page of Figure 11 of Lauer et
al. The blue line is the best-fit Nuker model, superimposed on the
(deconvolved) Nuker surface brightness profile (as shown in Figure 11
of Lauer et al.). The red curve shows the best-fit, PSF-convolved
ACSVCS model superimposed on the ACSVCS data. If a nucleus was fitted
to the data (as in the case of NGC4458 = VCC1146), the corresponding
PSF-convolved King profile is shown as a dotted line, while the
S\'ersic model for the galaxy is shown by the dashed line. The radius
represents the major axis radius for the Nuker data, and the geometric
mean radius for the ACSVCS data. Surface brightnesses refer to the
$V$-band for the Nuker data, and the $g$-band for the ACSVCS data. The
vertical line is drawn at 0\Sec05 (the size of an ACS/WFC pixel. The
WFPC2/PC and WFPC1/PC pixel scales are 0\Sec045 and 0\Sec043
respectively).  }
\end{figure}

\addtocounter{figure}{-1}
\addtocounter{subfigure}{1}
\begin{figure}
\includegraphics[scale=0.7]{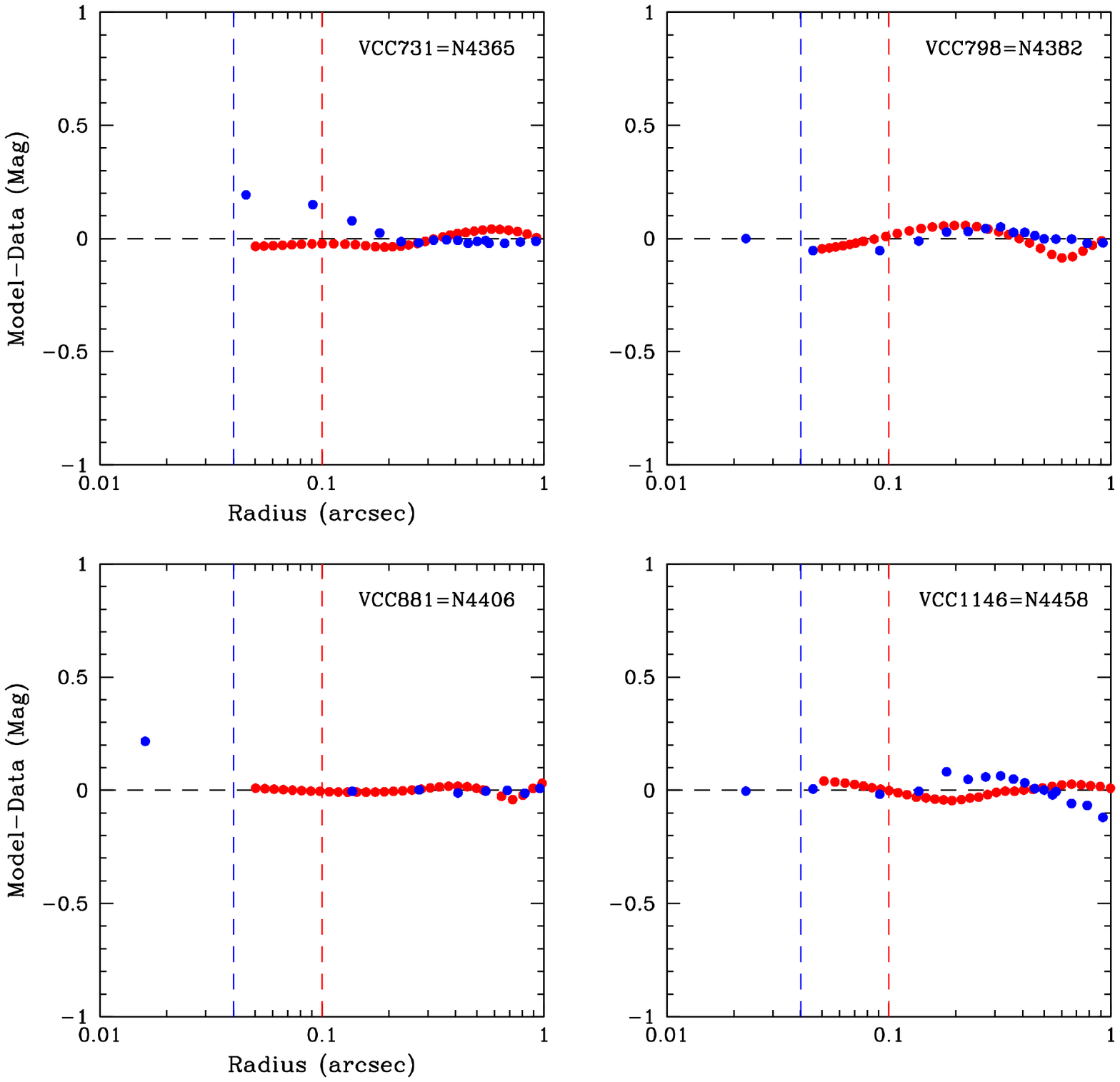}
\caption{Surface brightness residuals (model -- data) for the Nuker
model fits (blue) and the ACSVCS fits (red) for the galaxies shown in
Figure 4a. The red vertical line is drawn at 0\Sec1 (i.e., the radius
at which the ACSVCS slope was measured), while the vertical blue line
shows the resolution limit claimed by Lauer et al.\ for the
deconvolved WFPC1/WFPC2 data, where the slope was calculated.  }
\end{figure}

We now turn our attention to the quality of the model fits at small
radii ($r \lesssim$ 1\sec): i.e., the region of interest when
measuring the inner profile slope, $\gamma$. In Figures 4b to 9b, we
show residuals (model--data) in the innermost 1\sec, from which the
quality of the Nuker (in blue) and ACSVCS core-S\'ersic or S\'ersic
(in red) fits can be judged.  In all cases, the data and residuals are
plotted only up to the radius which was deemed reliable based on the
images (0\Sec05 in the case of the ACSVCS data).

Let us compare the fits on a galaxy by galaxy basis. Figures 4a and 4b
(corresponding to the first four panels of Figure 11 of Lauer et al.)
show that, in the case of NGC4365 (VCC731), the ACSVCS core-S\'ersic
model used by F06 provides a better representation of the data at
small radii than the Nuker model. For NGC4382 (VCC798) and NGC4406
(VCC881), the quality of the Nuker and core-S\'ersic model fits is
comparable on small scales; on larger scales, the Nuker model {\it
always} fares worse than the S\'ersic model.  NGC4458 (VCC1146) is
classified by F06 as a nucleated galaxy, and will be discussed in
detail in~\S 4.

Figures 5a to 5b  (corresponding to the second page of Figure 11 of
Lauer et al.)  show that the ACSVCS core-S\'ersic model is a better
representation of the data for NGC4472 (VCC1226), while the Nuker
model works better than the ACSVCS core-S\'ersic model for NGC4473
(VCC1231). At first glance, the Nuker model is also a significantly
better representation of the data for NGC4478 (VCC1279), however, this
galaxy  deserves a closer inspection. As discussed by F06, NGC4478
hosts an edge-on,  1\sec~ blue stellar disk at its center. The
presence of the disk is clear from the images themselves as well as
from the isophotal analysis, but it really jumps out from the
F475W-F850LP color image (see Figure 13 and notes in the Appendix of
F06). We will not speculate as to the reasons why the disk was missed
in the WFPC2 images analyzed by Lauer et al., but, in hindsight, the
irregularity in the deconvolved WFPC2 profile  (lower left panel of
the second page of Figure 11 of Lauer et al.) should have been a
giveaway (Lauer et al.\ note that NGC4478 has a ``small nucleus"). The
ACSVCS S\'ersic  model for this galaxy was not fitted within the inner
1\sec~to avoid contamination by the disk, thereby explaining the large
residuals in this region. It can, of course, be debated whether
extrapolating the ACSVCS model (which fits the profile between
1\sec~and 50\sec)  inward gives a faithful estimate of the intrinsic
profile slope at 0\Sec1 (the radius at which $\gamma$ was measured by
F06). What is certain is that the slope derived from the Nuker model
favored by Lauer et al. {\it does not}. The Nuker model was fitted
between 0\Sec1 and $\sim 10$\sec: by being forced to follow the disk's
profile for the first decade of this radial range, it is not fitting
the main body of the galaxy, but rather a component that is clearly
distinct in both morphology and stellar population. The slope
estimated by Lauer et al.\ is therefore certainly not a good estimate
of the $\gamma$ of the underlying galaxy. We note that, even in the
case of NGC4473 (for which, as noted above, the Nuker model provides a
better description of the data than the ACSVCS S\'ersic model), F06
pointed out the presence of a small, blue boxy feature in the $(g-z)$
color images, although in this case the evidence of a separate
component at the center is not as strong as for NGC4478.

\addtocounter{subfigure}{-1}
\begin{figure}
\includegraphics[scale=0.7]{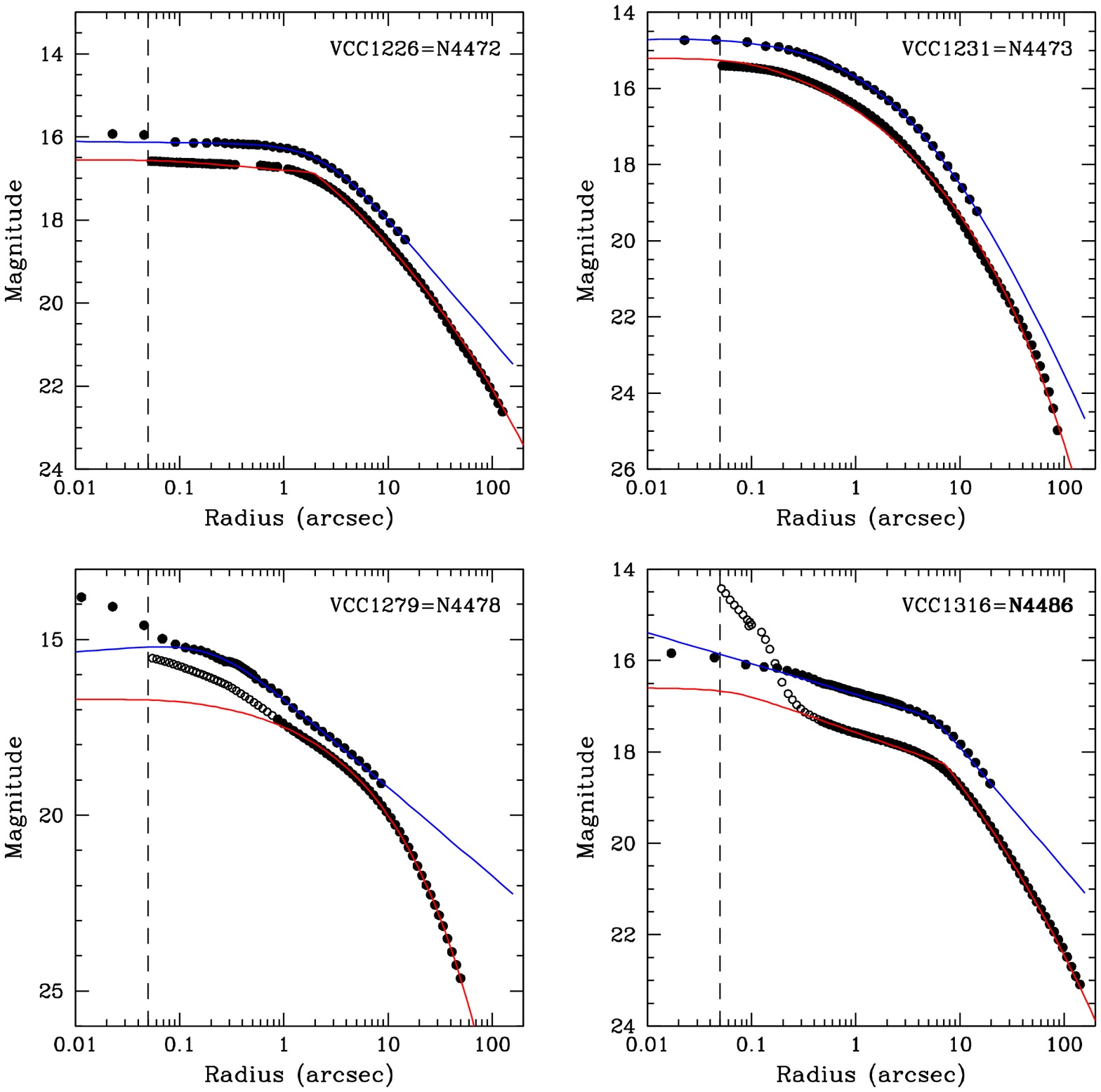}
\caption{The equivalent of the second page of Figure 11 of Lauer et
al.. The data used for the Nuker fit of NGC 4486 (M87, upper line) is
WFPC1/F785LP data from which the contribution of the bright
AGN was removed before deconvolution (Lauer et
al. 1992 -- the data needed to be shifted downwards by 1.35 mag to
match the model parameters tabulated in the Lauer et al. posting) --
in the original images, the AGN component dominates the galaxy
profile within the inner 1\sec. The core-S\'ersic model (red line) was
not fitted to the ACS data (lower curve) within 0\Sec45 (open symbols)
to avoid contamination of the AGN (the synchtron jet
was masked while fitting the isophotes).  In the case of NGC 4478, the
S\'ersic model (red line) was not fitted to the ACS data within 0\Sec8
(open symbols) to avoid contamination from the blue stellar disk
detected in the nuclear region of this galaxy (see discussion in the
text and F06).  }
\end{figure}

\addtocounter{figure}{-1}
\addtocounter{subfigure}{1}
\begin{figure}
\includegraphics[scale=0.7]{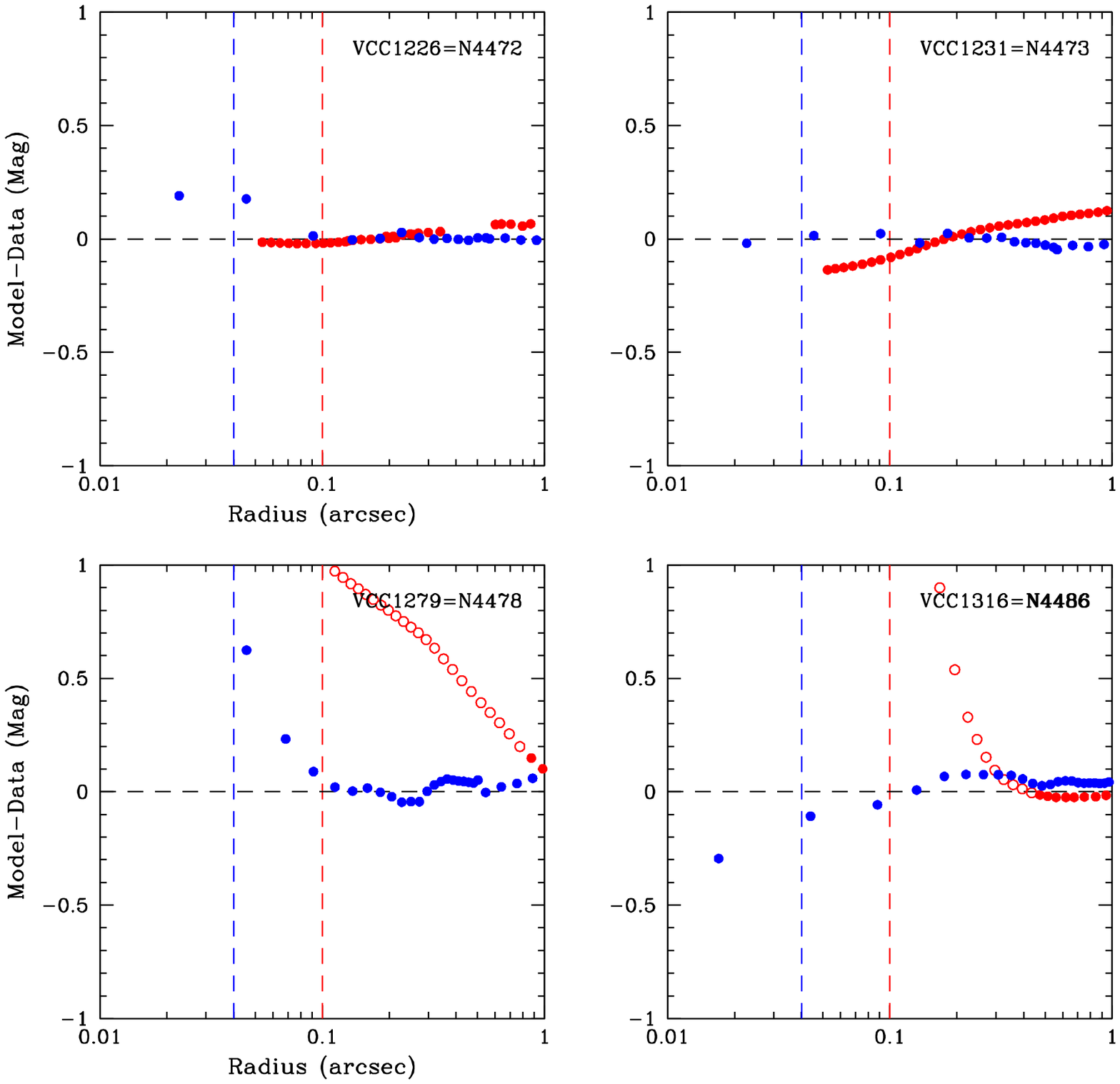}
\caption{Surface brightness residuals (model -- data) for the Nuker
model fits (blue) and the ACSVCS fits (red) for the galaxies shown in
Figure 5a. The ACSVCS models were not fitted to the data within the
radial range where residuals are shown as open circles.  }
\end{figure}

Finally, the galaxy in the bottom right panel of Figure 5a is NGC4486
(VCC1316 = M87). The data used by Lauer et al.\ to fit this galaxy are
WFPC1/F785LP (tranformed to Johnson $I$)  data from which the
contribution of the bright AGN and optical synchrotron jet
has been subtracted prior to deconvolution (from Lauer et al.\
1992). In the aberrated WFPC1 images, the AGN dominates
the inner 1\sec~  (see Figure 3 of Lauer et al.\ 1992). Using such
heavily processed data to fit a model down to 0.1 arcsec is risky to
say the least; using such data to argue for the superiority of the
Nuker fit over the core-S\'ersic fit from the ACSVCS is insupportable.
Our Figure 5a demonstrates that the ACSVCS model provides a good
representation of the  surface brightness profile between 0\Sec3 and
over 100\sec. Inside 0\Sec3, the profile is dominated by 
emission from the central AGN, and this region was excluded in the
ACSVCS fit for precisely this reason.

Moving on to Figures 6a, 6b, 7a and 7b, the ACSVCS and Nuker fits are
of comparable quality for NGC4486B (VCC1297)\footnotemark, NG4649
(VCC1978), NGC4621 (VCC1903) and NGC4434 (VCC1025), while the Nuker
model fits are somewhat better (on small scales) for NGC4552 (although
the Nuker model fit is not a good match to the profile at 0\Sec4,
where the Nuker model slope is measured), NGC4464 (VCC1178) and
NGC4660 (VCC2000).

\footnotetext{Note that in Figure 11 of Lauer et al., the
  galaxy at the top left is labeled as NGC4486 when in fact it is
  NGC4486B.}

In summary, there is no evidence from the fits shown in these figures
that, on small radial scales ($r \lesssim 1$\sec), the Nuker models
perform consistently better than the ACSVCS models. The Lauer et
al. claim to the contrary is driven by a misleading comparison of the
ACSVCS models to data taken with a different filter, instrument, and
described by a different PSF. In at least one case (NGC4478), Lauer et
al.\ failed to recognize and properly account for the existence of a
morphologically separate nuclear component; in this case, the Nuker
model fit adopted by Lauer et al.\ is certainly in error. On larger
scales ($r \gtrsim  10$\sec), the ACSVCS models are {\it always} a
better description of the data than the Nuker models.

\addtocounter{subfigure}{-1}
\begin{figure}
\includegraphics[scale=0.7]{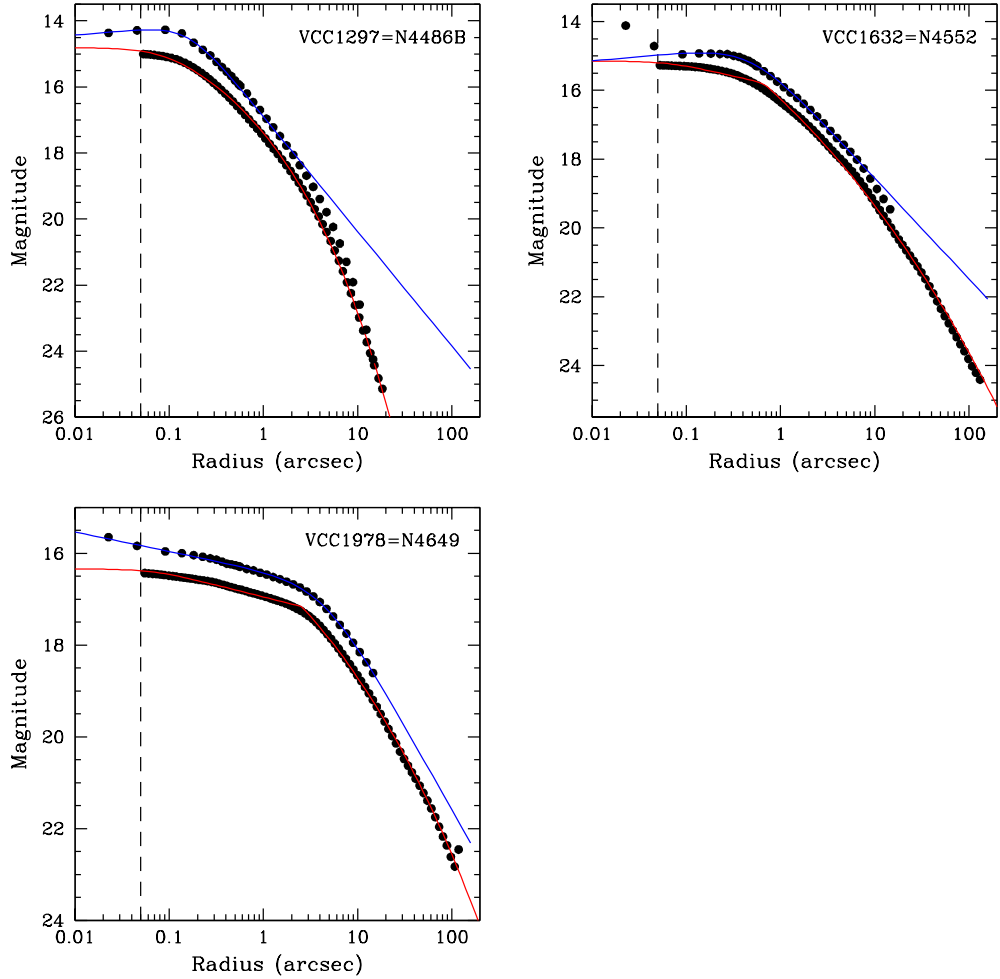}
\caption{The equivalent of the third page of Figure 11 of Lauer et
al. 
}
\end{figure}

\addtocounter{figure}{-1}
\addtocounter{subfigure}{1}
\begin{figure}
\includegraphics[scale=0.7]{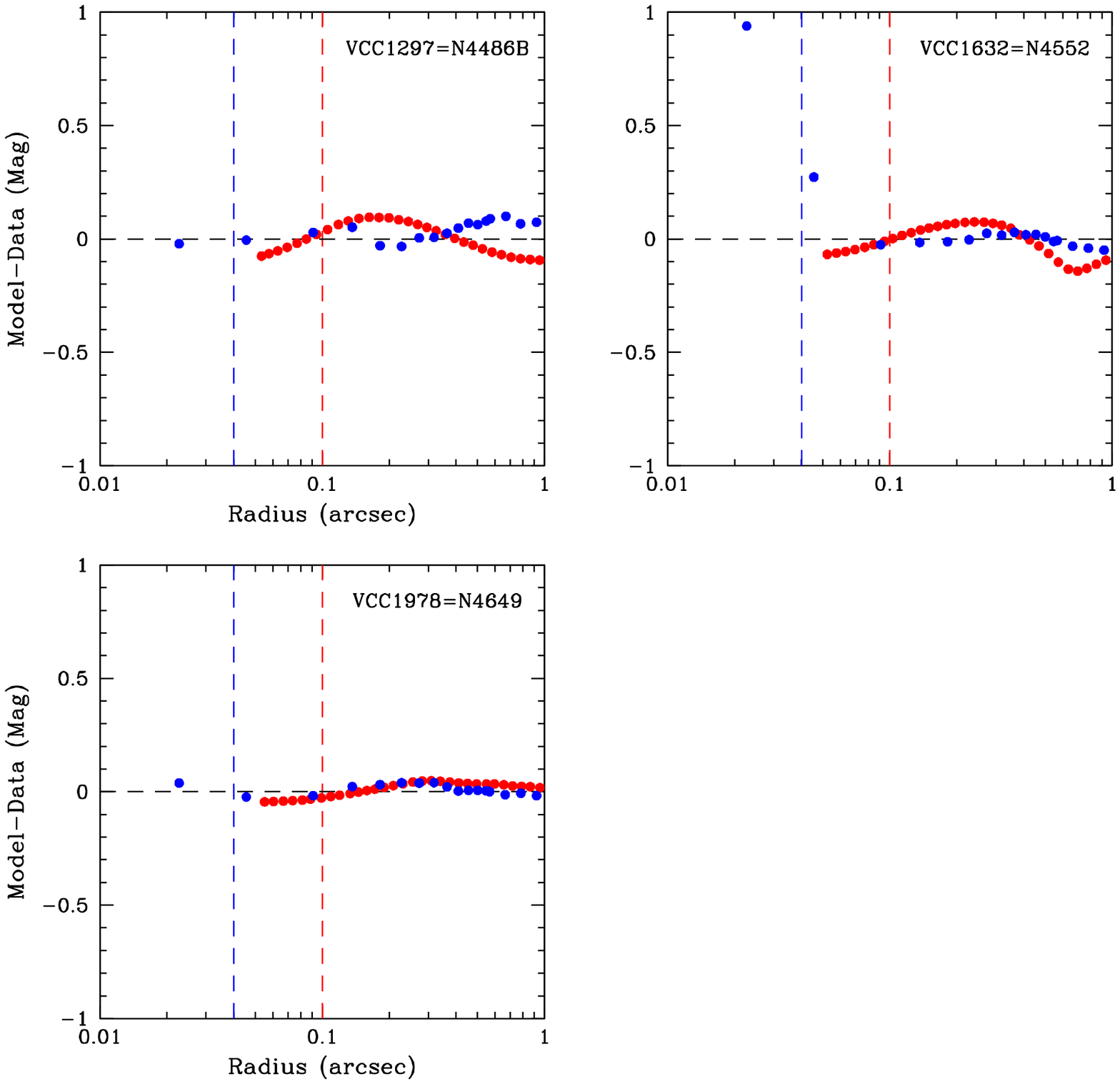}
\caption{Surface brightness residuals (model -- data) for the Nuker 
model fits (blue) and the ACSVCS fits (red) for the galaxies shown 
in Figure 6a.
}
\end{figure}

\addtocounter{subfigure}{-1}
\begin{figure}
\includegraphics[scale=0.7]{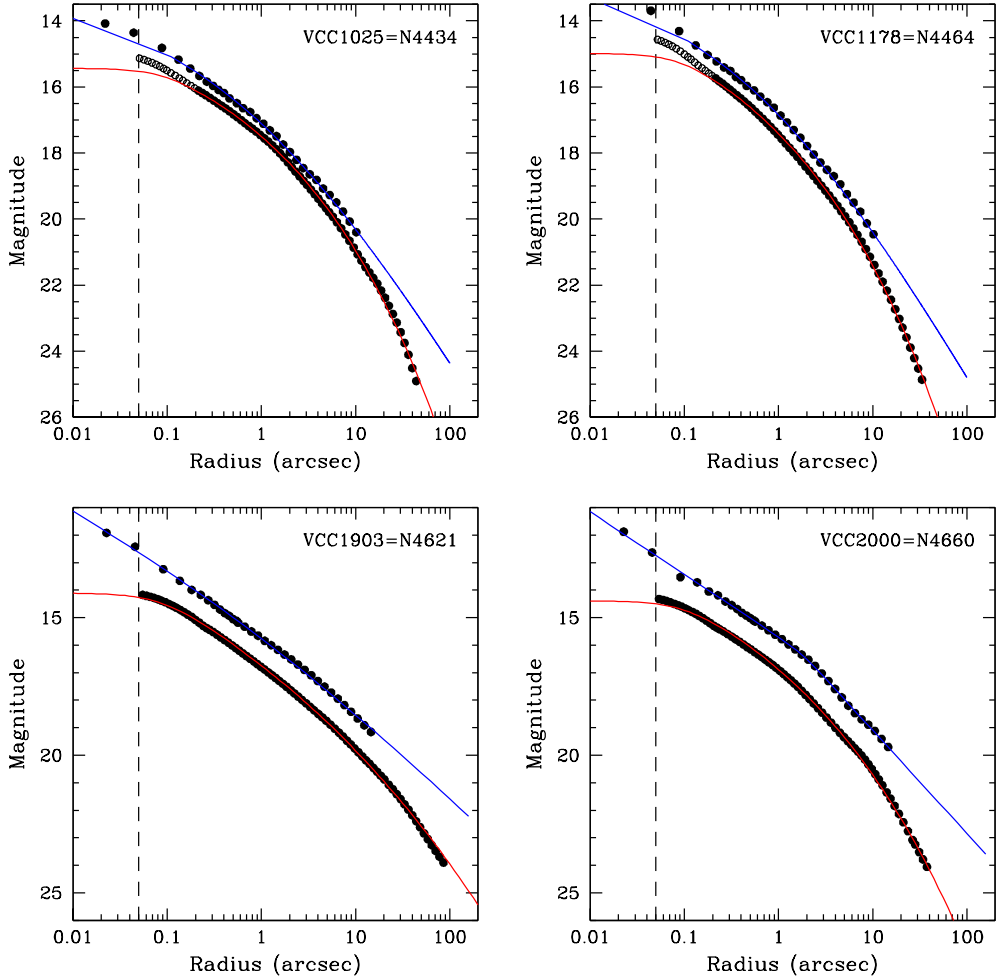}
\caption{The equivalent of Figure 12 of Lauer et al.. The ACSVCS
models (shown in red) were not fitted within the inner 0\Sec2 of NGC
4434 and NGC 4464 (shown as open symbols). A compact stellar nucleus
was identified within this region, although the small contrast between
the nucleus and the underlying galaxy prevented from deriving an
accurate King model fit. Note that the Nuker model favored by Lauer et
al. (blue line) does not provide a good fit to the profile within this
region. In the case of NGC 4621 and NGC 4660, we show the F850LP,
rather than F475W ACS profiles -- the latter were saturated within the
inner 0\Sec25, a region that was therefore omitted in fitting the
F475W profiles. The F850LP profiles and models have been shifted
downwards by 2 mag for clarity.  }
\end{figure}

\addtocounter{figure}{-1}
\addtocounter{subfigure}{1}
\begin{figure}
\includegraphics[scale=0.7]{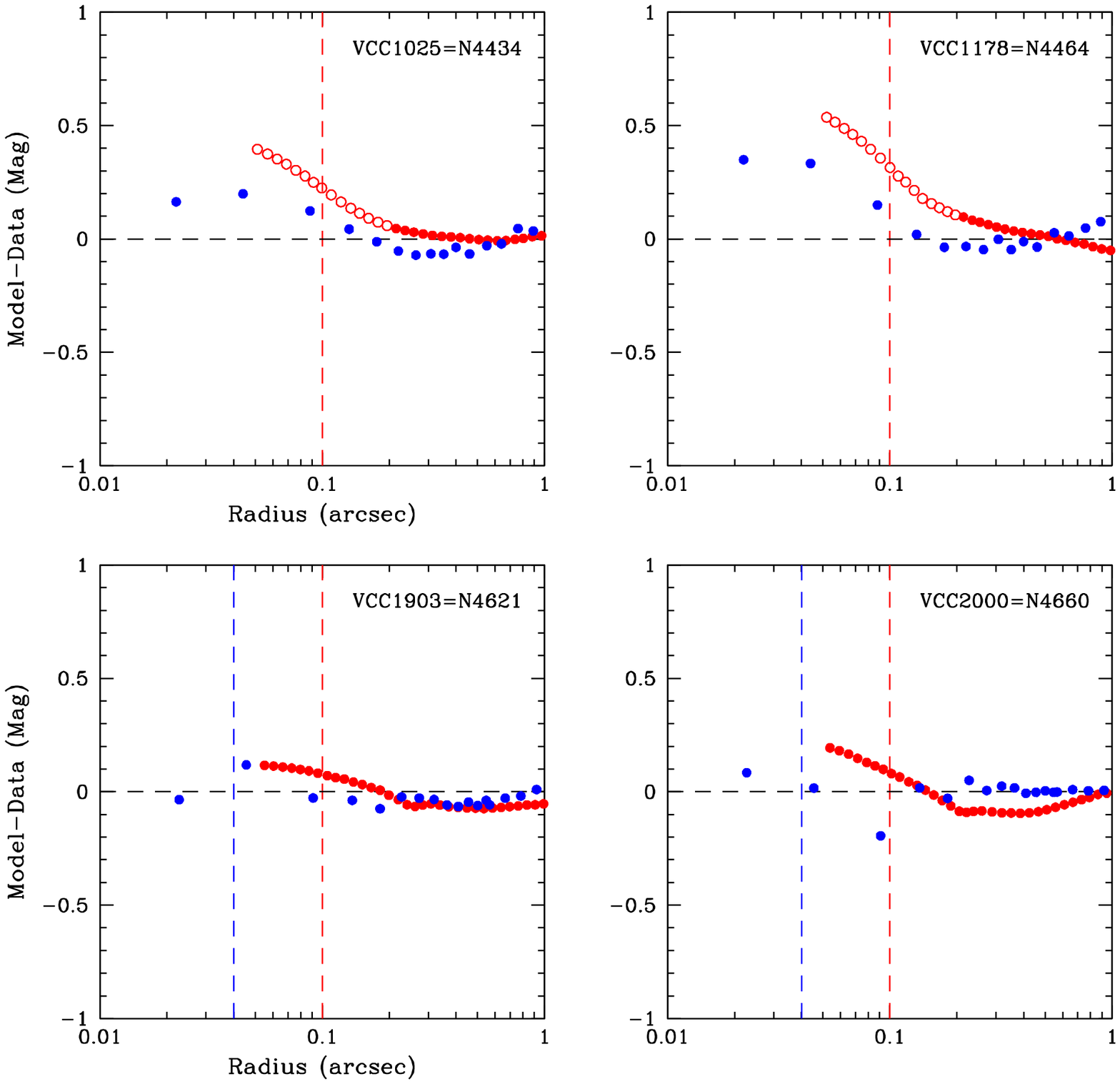}
\caption{Surface brightness residuals (model -- data) for the Nuker
model fits (blue) and the ACSVCS fits (red) for the galaxies shown in
Figure 7a. The ACSVCS models were not fitted to data within the radial
range in which residuals are plotted as open symbols.  }
\end{figure}

\section{Identification of Stellar Nuclei}

\addtocounter{subfigure}{-1}
\begin{figure}
\includegraphics[scale=0.7]{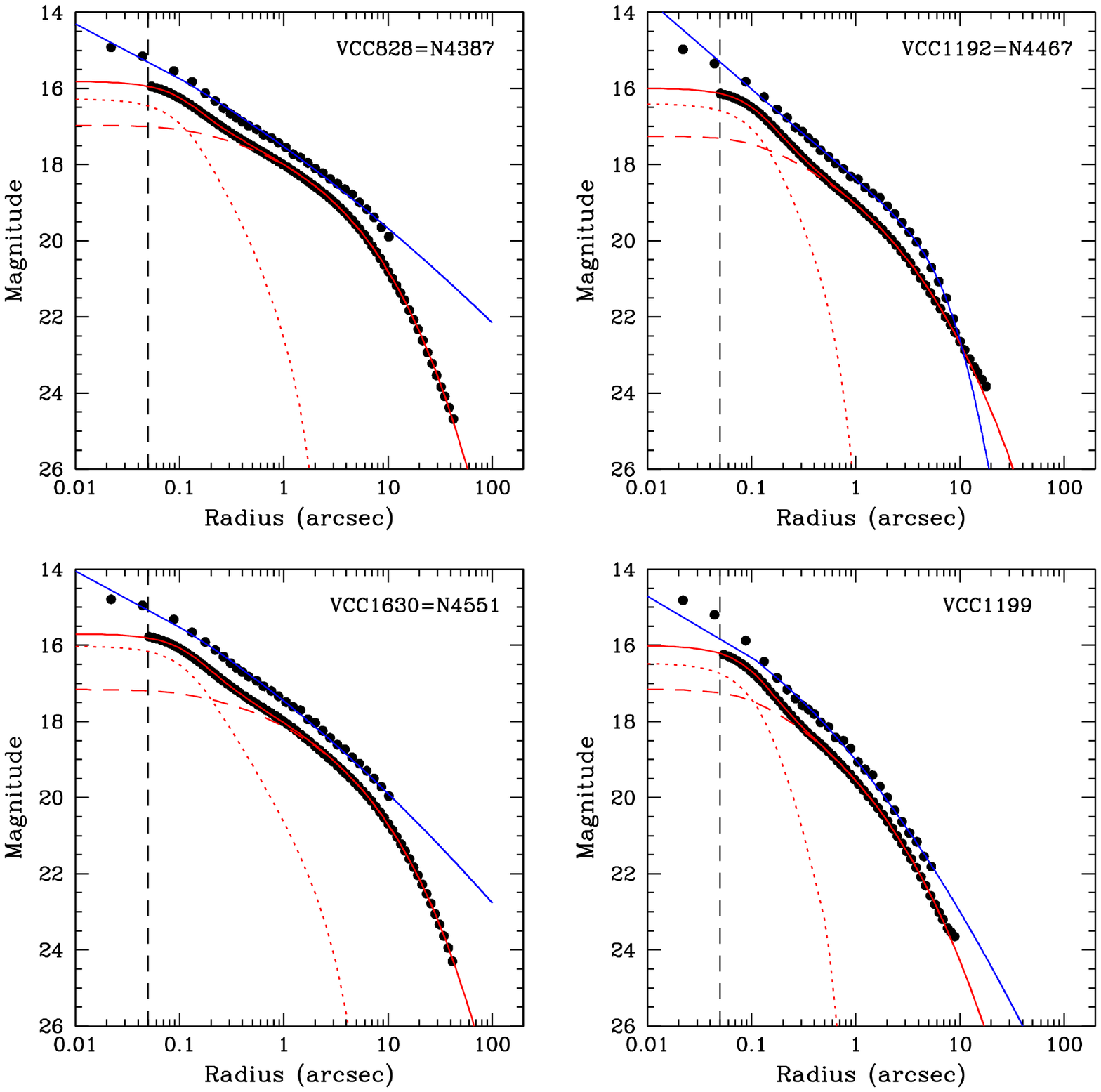}
\caption{The equivalent of the first page of Figure 13 of Lauer et
al. For all galaxies shown in this figure, deconvolved WFPC1 data 
were used by Lauer et al.\ to fit Nuker models.
}
\end{figure}

\addtocounter{figure}{-1}
\addtocounter{subfigure}{1}
\begin{figure}
\includegraphics[scale=0.7]{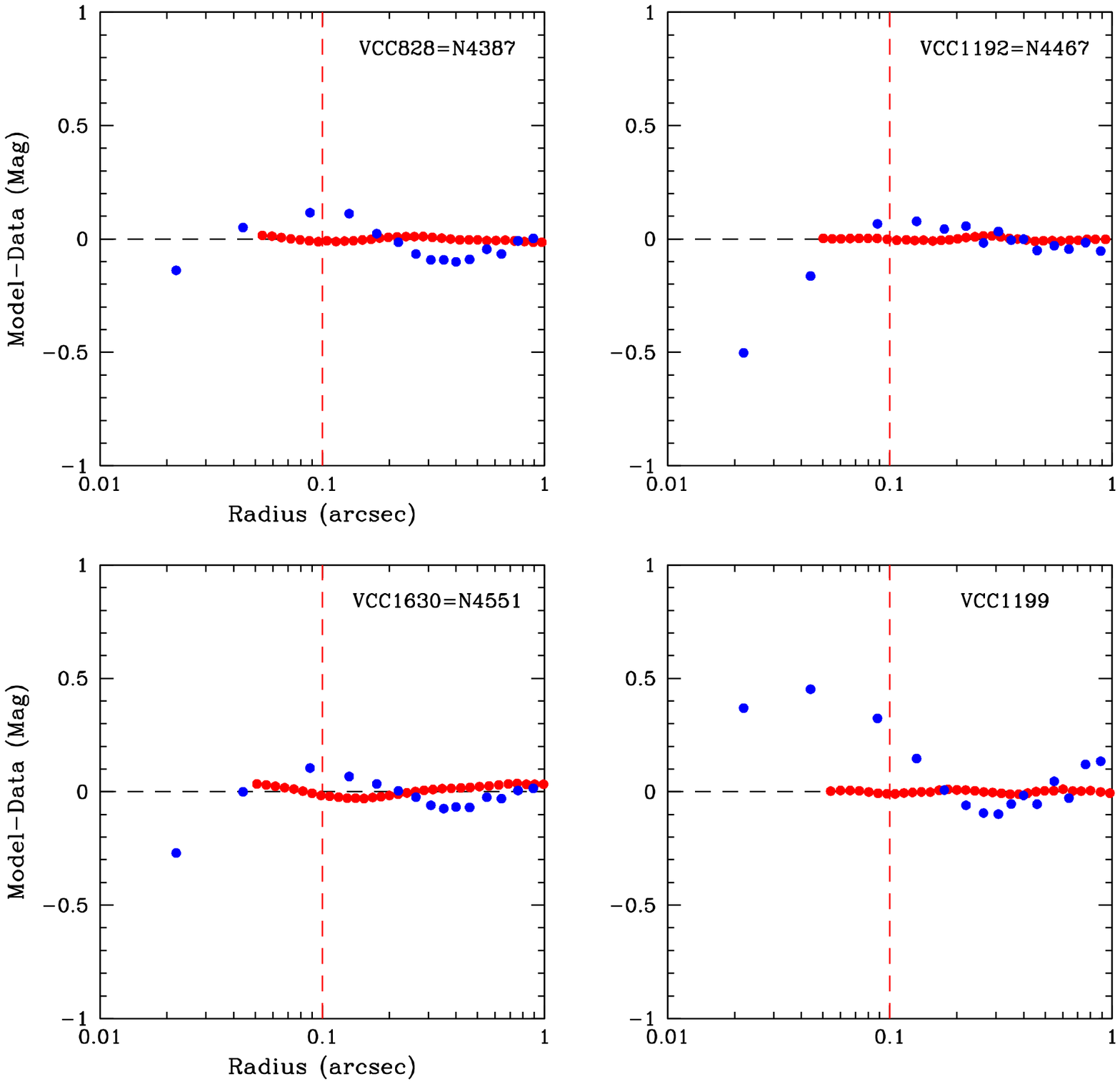}
\caption{Surface brightness residuals (model -- data) for the Nuker 
model fits (blue) and the ACSVCS fits (red) for the galaxies shown 
in Figure 8a.
}
\end{figure}

\addtocounter{subfigure}{-1}
\begin{figure}
\includegraphics[scale=0.7]{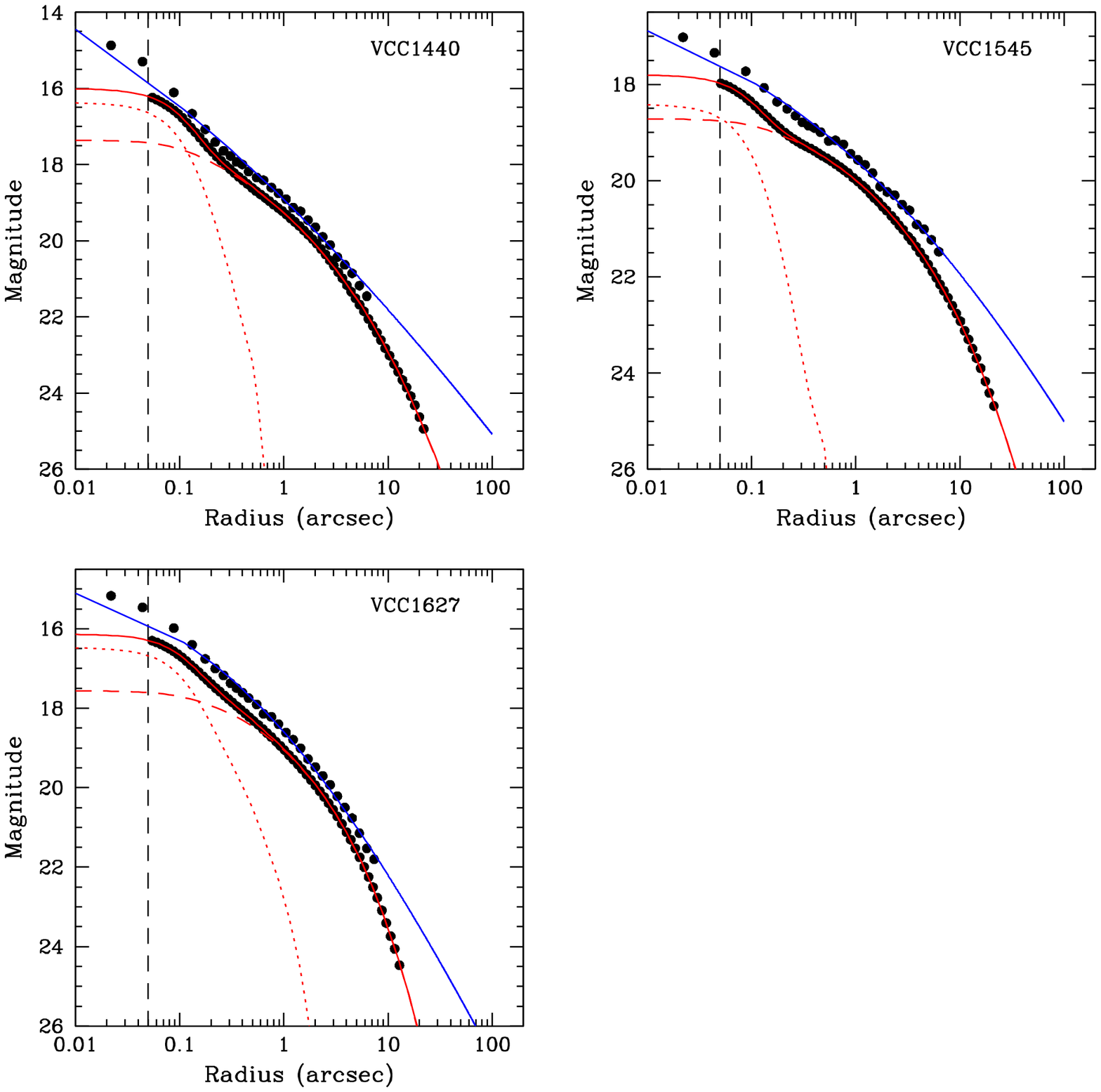}
\caption{The equivalent of the second page of Figure 13 of Lauer et
al. For all galaxies shown in this figure, deconvolved WFPC1 data 
were used by Lauer et al.\ to fit Nuker models.
}
\end{figure}

\addtocounter{figure}{-1}
\addtocounter{subfigure}{1}
\begin{figure}
\includegraphics[scale=0.7]{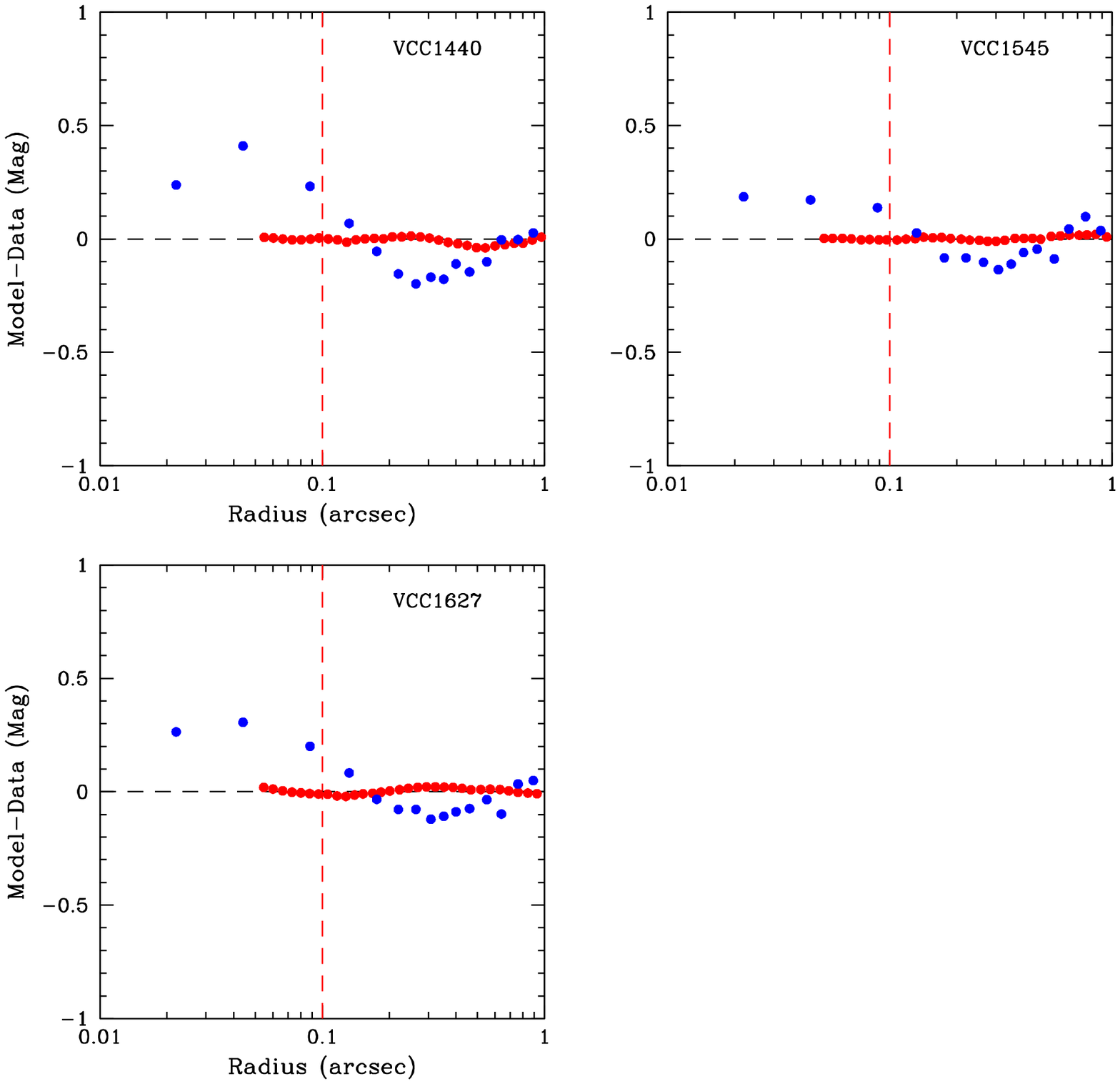}
\caption{Surface brightness residuals (model -- data) for the Nuker 
model fits (blue) and the ACSVCS fits (red) for the galaxies shown 
in Figure 9a.
}
\end{figure}

The second criticism advanced by Lauer et al.\ concerns  the
definition of stellar nuclei. F06 and C06 identified nuclei from a
variety of diagnostics, including sudden upturns in the surface
brightness profiles and color changes in the $(g-z)$ color images. In
clearly nucleated galaxies, nuclei were fitted by King models added to
the S\'ersic model representing the main body of the galaxy. Lauer et
al.\ argue that the nuclei identified by F06 and C06, in most of the
ACSVCS galaxies, are spuriously introduced to fill in the gap left
over by the fact that the fitted S\'ersic models underestimate the
profile in the inner region: i.e., they state ``... excursions of the
data above the S\'ersic model are declared to be separate nuclear
components, rather than as a simple failure of the model. The VCS
nuclei effectively absorb the central flux left over from the S\'ersic
fits". Referring to the galaxies shown in their Figure 13 (classified
as nucleated by F06 and C06), Lauer et al.\ comment that ``there are
no strong upward breaks in any of the galaxies discussed in this
section [shown in Figure 13] that would make detection of a nucleus
unambiguous''. Yet this statement is in plain contradiction with a
statement made only a few paragraphs earlier:  ``Lauer et al.\ (1995)
identified nuclei in all of these galaxies as well [i.e. the galaxies
shown in Figure 13], albeit ones of markedly lower luminosity and
extent than those presented by C\^ot\'e et al.\ (2006)."

Therefore, despite their claim to the contrary, it is not the {\it
existence} of a nucleus that is called into question, but the {\it
definition}. F06 and C06 define nuclei as excesses with  respect to
S\'ersic models, while Lauer et al.\ ``identify nuclei by looking for
upturns above a power-law cusp as $r \rightarrow 0$". Thus, both Lauer
et al.\ and F06 and C06 define nuclei as upturns relative to the
inward extrapolation of the model that best fits the outer parts of
the profile. In view of this, it is difficult to understand the Lauer
et al.\ dismissal of the ACSVCS approach (``takes it as an a priori
assumption rather than as a hypothesis that the envelopes of galaxies,
which is where the S\'ersic models are fitted, can be used to deduce
the structure of the central profile at small radii.") given that
Lauer et al.\ choose a Nuker  model to make precisely this same
decision.  The Lauer et al.\ criticism is even more puzzling when one
realizes that the ``outer envelope" used by F06 and C06 to fit
S\'ersic models is, in fact, nearly the full extent of the galaxy
(i.e., a region between a few 0\Sec1 to $20$\sec$ - 100$\sec). The
Lauer et al.\ Nuker fits, on the other hand, are typically shown
between 0\Sec1 and a few arcsec (see their Figure 13). In short, Lauer
et al.\ appear to argue that having surface brightness data with the
maximum possible radial coverage, and using it to find the models that
best fit the global profile, is a detriment rather than an advantage.

At this point, the only relevant question is which of a S\'ersic or a
Nuker model does a better job at fitting the profiles in the region
beyond the nucleus. Lauer et al.\ claim that a Nuker model does: ``It
is evident from examining the profiles fits in Figure 12 [typo: should
be  Figure 13] that the Nuker laws can accurately describe the ACS
profiles of these galaxies into small radii." The veracity of this
claim is tested in Figures 8a, 8b, 9a and 9b, where we show the Nuker
model against the data to which it was originally fit, and the
S\'ersic and King (both combined and separate) models against the
ACSVCS data (F06, C06). As was the case for Figures 4$-$6, it is worth
noticing once again the wider radial coverage of the ACSVCS data, but
also the fact that the data used by Lauer et al.\ are considerably
noisier than the ACSVCS data. The Lauer et al.\ Nuker model fits to
these galaxies use deconvolved, pre-refurbishment WFPC1 data and the
intrinsically low S/N of the data is further amplified in their
deconvolution process. At any rate, the inescapable conclusion from
the residuals shown in these figures is that, in every case, and in
stark contrast to the claim of Lauer et al., the Nuker model of Lauer
et al.\ {\it does not} provide a good description of the data at {\it
any} radius; in the innermost region in particular, the residuals
always show a characteristic {\tt S}-shaped signature.

The ACSVCS S\'ersic+King model, on the other hand, provides an
excellent description of the data. The reader might wonder why this is
not the impression one is left with when looking at Figure 13 of the
Lauer et al. posting. There are three reasons.  First, Lauer et al.\
misrepresent the ACSVCS analysis by showing only the S\'ersic models,
and omitting the contribution of the nucleus that is essential in the
ACSVCS description of these galaxies. Second, they plot pre-convolved
S\'ersic models against deconvolved data, an inappropriate comparison
as argued earlier. Finally, for both S\'ersic and Nuker models, they
plot a very limited radial extent (typically the inner 3-5\sec, but as
little as 2\sec~ for VCC 1199) and show no residuals, making it
impossible for the reader to appreciate the global trends in the
surface brightness profiles and the overall quality of the
fits. Within the restricted radial range plotted by Lauer et al.\,  it
is indeed true that an upturn in the surface brightness profile is not
``unambiguous". But when the full extent of the profile is plotted, as
shown in our Figures 8 and 9, the upturns are unmistakable. Indeed,
because of the limited radial extent of the data to which the Nuker
model was fitted, Lauer et al.\ failed to properly characterize the
nuclei, and missed the existence of the nuclear scaling relations
discussed in C06, Ferrarese et al.\ (2006b) and Wehner \& Harris
(2006).  Contemporaneous work in spiral galaxies suggests that similar
scaling relations are followed by the nuclei in these environments as
well (e.g., B\"oker et~al. 2002, 2004; Rossa et al. 2006; Seth et
al. 2006).

Finally, we make one last comment regarding NGC4458 = VCC1146 (bottom
right panel of Figures 4a and 4b). This galaxy also shows a central
upturn, which F06 and C06 take as indication that a central component
is present. Figure 14 of Lauer et al., by plotting only the S\'ersic
ACSVCS model and not the fitted nuclear component, and by showing only
the WFPC2 data (not the ACSVCS data), is  misleading. However, it is
interesting to note that, while the Nuker fit to this galaxy does an
excellent job in the inner 0\Sec1, the fit at larger radii shows the
typical ``{\tt S}-shaped" residuals that are generally seen when a
nucleus is present. Lauer et al., in their attempt to fit the
innermost region of this galaxy, are in fact fitting a Nuker model to
this nuclear component, and in the process producing a very poor fit
to the main body of the  galaxy --- the very component their model is
intended to represent.

The question remains, of course, whether any model fitted to the data
beyond the central nucleus and extrapolated inwards, can give a
truthful estimate of the slope of the profile of the galaxy underlying
the nucleus at that radius. This is indeed an interesting question,
and one to which there can be no secure answer. We will point out,
however, that a S\'ersic model does a  remarkably good job at fitting
the profiles of faint non-nucleated galaxies up to the innermost
radius  corresponding to the resolution element of the ACS
(e.g. VCC1049, VCC1833, VCC9, VCC1499, VCC1857, VCC1948, see Figure
103 of F06). In nucleated galaxies, it fits the profiles between a few
0\Sec1 and several tens of arcsec. Based on this, the assumption that
it might also describe the surface brightness profile underlying the
nucleus ($r \lesssim$ a few 0\Sec1) does not appear outlandish. Lauer
et al.\ Nuker models, on the other hand, are often a compromise fit
between the main body of the galaxy and the nucleus, which is not
properly recognized and accounted for because of the generally low S/N
of the data (compounded by the deconvolution procedure), the limited
radial extent of the surface brightness profiles, and the lack of
color information for many of their galaxies, from which the presence
or absence of a separate nuclear component could be ascertained.

\section{Discussion and Conclusions}

\renewcommand{\thefigure}{\arabic{figure}}

The parameterization of the surface brightness profiles of early-type
galaxies has been long used to characterize scaling relations and
differences/commonalities between these systems. In Ferrarese et
al. (2006, F06), and C\^ot\'e et al. (2006, C06), the surface
brightness profiles of 100 early-type galaxies in the Virgo cluster,
each observed with the ACS/WFC on board HST, were fitted using
core-S\'ersic, S\'ersic, or S\'ersic+King models. In the previous
sections, we have compared the quality of such fits with that provided
by ``Nuker" models, used in a recent astro-ph posting by Lauer et
al. to describe WFPC1 or WFPC2 data for a sample of galaxies in common
with the F06 and C06 sample. We argue that the Nuker model, while
being inadequate at describing the surface brightness profiles of
early-type galaxies on kiloparsec scales,  even on small scales does
not provide a characterization of the profiles that is superior to
that provided by the models used in the ACSVCS analysis. Indeed, we
have shown that the limited radial extent and low S/N for much of the
Virgo data used by Lauer et al., was responsible for the fact that
these authors failed to properly identify and characterize stellar
nuclei and that, as a consequence, their Nuker models are forced to
partially fit the profile of the nucleus, rather than that of the
underlying galaxy.  For the rest of this contribution, we point out a
few other issues of concern with the analysis presented in the Lauer
et al.\ astro-ph posting.

We start by examining the working definition of $\gamma$ used Lauer et
al., as the logarithmic slope measured at the resolution limit of the
instrument. Given that the Lauer et al.\ compilation was observed with
a variety of instrument/filter combinations, the angular scale at
which $\gamma$ is measured is different for each sample. Furthermore,
given the enormous range in distance (a factor 100) spanned by their
galaxies, $\gamma$ is measured at {\it very} different physical scales
in each galaxy.

According to Lauer et al., the instrument resolution limit is  0\Sec04
for WFPC2 (except for some unspecified galaxies for which drizzled
images exist, and for which the limit is taken to be 0\Sec02) and
0\Sec1 for WFPC1, NIC2 and NIC3.\footnote{Note that the pixel scales
of the instruments are 0\Sec043 (WFPC1/PC), 0\Sec045 (WFPC2/PC),
0\Sec075 (NIC2) and 0\Sec2 (NIC3). The resolution limits adopted by
Lauer et al. appear rather optimistic, given that they are a factor
2.25 smaller than a pixel for the drizzled WFPC2 images, that the FWHM
of the NIC2 and NIC3 PSFs are 0\Sec14 and 0\Sec22 respectively, and
that the NICMOS images of Ravindranath et al.\ were not deconvolved.}
This lack of uniformity is an obvious concern: Lauer et al.\
themselves note that when $\gamma$ is recalculated at 0\Sec04 using
the Nuker model parameters of Rest et al. (2001) --- who themselves
judged 0\Sec1 to be a more appropriate choice based on the analysis of
their own data --- in 12 cases (25\% of the sample) $\gamma$ changes
by a sufficiently large amount to move the galaxy classification
within the ``core -- intermediate -- power-law'' scheme favored by
Rest et al. (2001) -- with galaxies preferentially been moved out of
the ``intermediate" class into the ``core" class\footnotemark
\footnotetext{Seven galaxies, classified by Rest et al. as
intermediate cases are reclassified by Lauer et al. as core galaxies,
while four galaxies classified by Rest et al. as power-law, are
reclassified by Lauer et al. as intermediate. One galaxy, classified
by Rest et al. as power-law, is reclassified by Lauer et al. as core.}
. One is therefore left to wonder how the bimodal distribution of
$\gamma$ values found by Lauer et al.\ would be affected if the
galaxies observed with WFPC1, NIC2 and NIC3 had instead been observed
with WPFC2, in which case $\gamma$ would have been measured at a
radius smaller by a factor of $\sim$ 2.5--5.

This critical issue is further clouded by the fact that the Lauer et
al.\ slope is {\it not always} measured at the resolution limit of the
instrument. As Lauer et al.\ state, ``the limits shown are those we
have adopted to avoid central nuclei, and so on, and when larger than
the general resolution limits presented in \S2 are always the scale at
which we measured $\gamma$". Therefore, for galaxies that are believed
to be nucleated, the slope is {\it measured at the innermost radius
that is believed not to be affected by the nuclear component}. Lauer
et al.\ do not actually tabulate the radius at which $\gamma$ is
measured in each individual galaxy, but for NGC4365, for instance,
their Figure 11 shows that $\gamma$ is measured at a radius that is a
factor 10 larger than the instrumental resolution limit. Therefore,
even if one could find a good justification as to why $\gamma$ should
be measured at the resolution limit of the instrument (which we
cannot, and certainly not for for a sample of galaxies lying at
different distances and observed with instruments for which this limit
varies by a factor five), Lauer et al.\ violate their own rule in
their analysis.

Indeed, casting the radius at which $\gamma$ is measured in terms of
an angular scale has very little sense when dealing with a sample of
galaxies which span a factor 100~in distance (as already pointed out
by Graham et al. 2003). If the core/power-law bimodality has a
physical origin, it would seem more appropriate to measure the slope
at either the same physical, rather than angular radius, or at least
at a radius corresponding to a constant fraction of some
characteristic scale radius in every galaxy  (such as the effective
radius of the galaxy). The only condition here is that such radius
must be chosen to be smaller than the break radius observed for the
brightest galaxies.

To summarize, the main points from the preceding sections are:

\begin{itemize}

\item The ACSVCS sample is superior to the sample compiled by Lauer et
al.\ in terms of:

\begin{enumerate}

\item Spatial resolution (better by an average factor of $\sim$
three);

\item Homogeneity. All ACSVCS galaxies are observed with the same
instrument/filter combination, while the Lauer et al.\ is an
inhomogeneous compilation of samples observed with four different
instruments and in four different bandpasses;

\item Sample selection. All ACSVCS galaxies belong to the Virgo
Cluster and uniformly cover the luminosity function of early-type
galaxies. The Lauer et al.\ sample includes objects observed as part
of five different projects so its biases and completeness are not
easily characterizable;

\item Availability of two bands for all galaxies. The color
information is an important factor in assessing the presence of
stellar nuclei and separate morphological components; these cannot be
easily recognized from single-band data such as those used by Lauer et
al.;

\item Spatial coverage. The radial extent of the brightness profiles
in common between the two studies is typically greater by an order of
magnitude for the ACSVCS sample.

\end{enumerate}

\item In non-nucleated galaxies, Lauer et al. claim that the
core-S\'ersic and S\'ersic models used to fit the ACSVCS data do not
provide a good characterization of the profile. This claim is based on
a  comparison of the ACSVCS models to deconvolved WFPC2 or WFPC1 data
taken with a different filter (i.e., the data used to fit Nuker
models). This comparison is fundamentally inappropriate since it does
not account for the PSF mismatch between the ACSVCS models and the
deconvolved images. When the ACSVCS models are compared to the ACSVCS
data, their ability to reproduce the profiles at small radii is
comparable to that of the Nuker model fitted to deconvolved WFPC1 or
WFPC2 data. At the same time, the ACSVCS models provide good fits at
large radii, whereas the Nuker models fail dramatically on such scales.

\item In nucleated galaxies, Nuker models are poor fits to the inner
profiles despite the claim of Lauer et al. to the contrary.  ACSVCS
models for these galaxies, including a King component that describes
the nuclei, do an excellent job of matching the profile. The low S/N
and limited radial extent of the WFPC1 data used by Lauer et al.\ to
fit Nuker models to these galaxies prevented them from identifying the
global trends in the profiles, which are essential when judging the
extent of contamination by central components. As a result, the fitted
Nuker profiles are a compromise between the nucleus and the underlying
galaxy, fitting neither particularly well, as is shown by the clear
{\tt S}-shaped signature imprinted in the Nuker model residuals.

\end{itemize}

In short, on small scales, a Nuker model does not provide a better
description of the data than a S\'ersic or core-S\'ersic model. On
kiloparsec scales, a Nuker model is unable to follow the continuous
curvature that characterizes galaxy profiles. Indeed, Graham et
al. (2003) argued that the curvature of the profiles on large scales
undermines the use of a Nuker model by making the fits sensitive to
the radial extent covered by the data, to the point that the physical
interpretation of the model parameters is compromised.

Core-S\'ersic, S\'ersic, or S\'ersic+King models appear better suited
to describe the surface brightness profiles of early-type galaxies
from parsec to kiloparsec scales. The fitted parameters do not appear
to suffer  from significant biases, and the models use a relatively
small number of free parameters (three for non-nucleated galaxies, and
six when a nucleus is present). As more and better surface brightness
data become available, it is possible --- and indeed, likely --- that
shortcomings of these models will also begin to emerge, and that a
newer and better parameterization will be required. Until then, a
S\'ersic/core-S\'ersic parametrization should be preferred to that
offered by a Nuker model.

\end{document}